\documentclass[12pt,letterpaper]{article}
\pdfoutput=1

\usepackage[T1]{fontenc}
\usepackage[utf8]{inputenc}
\usepackage{lmodern}
\usepackage{amsmath,amssymb,amsthm,mathtools,bm,mathrsfs}
\usepackage{microtype}
\usepackage[margin=1in]{geometry}
\usepackage[numbers,sort&compress]{natbib}
\usepackage{enumitem}
\usepackage[dvipsnames]{xcolor}
\usepackage{hyperref}

\hypersetup{
  colorlinks=true,
  citecolor=MidnightBlue,
  linkcolor=MidnightBlue,
  urlcolor=MidnightBlue,
  pdfauthor={Christian Ferko, Cian Luke Martin, Pranat Sharma},
  pdftitle={Auxiliary Field Deformations of the Lambda Model}
}

\numberwithin{equation}{section}
\allowdisplaybreaks

\theoremstyle{remark}

\DeclareMathOperator{\Ad}{Ad}
\DeclareMathOperator{\ad}{ad}
\newcommand{\tr}{\operatorname{tr}}
\newcommand{\dd}{\mathrm{d}}
\newcommand{\id}{\mathbf{1}}

\makeatletter
\newcommand*\bigcdot{\mathpalette\bigcdot@{.5}}
\newcommand*\bigcdot@[2]{\mathbin{\vcenter{\hbox{\scalebox{#2}{$\m@th#1\bullet$}}}}}
\makeatother

\renewcommand{\doteq}{\stackrel{\bigcdot}{=}}

\newcommand{\Lax}{\mathfrak{L}}

\begin{document}

\hypersetup{pageanchor=false}
\begin{titlepage}
\begin{flushright}
\today
\end{flushright}
\vspace{5mm}

\begin{center}
{\Large\bfseries Auxiliary Field Deformations of the Lambda Model}
\end{center}

\begin{center}
{\bfseries
Christian Ferko${}^{a,b}$,
Cian Luke Martin${}^{c}$,
and Pranat Sharma${}^{d}$
}\\
\vspace{5mm}

\footnotesize
${}^{a}$\textit{Department of Physics, Northeastern University, Boston, MA 02115, USA}
\\[2mm]
${}^{b}$\textit{The NSF Institute for Artificial Intelligence and Fundamental Interactions}
\\[2mm]
${}^{c}$\textit{Department of Physics and Astronomy, Stony Brook University,\\
Stony Brook, NY 11794-3840, USA}
\\[2mm]
${}^{d}$\textit{Brooklyn College of the CUNY, Brooklyn, NY 11210, USA}
\\[3mm]
\texttt{c.ferko@northeastern.edu,
cian.lukemartin@stonybrook.edu, 
pranat.sharma01@brooklyn.cuny.edu}
\end{center}

\vspace{3mm}
\begin{abstract}

\noindent Many examples of integrable sigma models admit families of integrable deformations that involve coupling the undeformed ``seed'' theory to auxiliary fields with algebraic equations of motion. In this work, we apply this paradigm to the case where the seed theory is the lambda model, which interpolates between the Wess-Zumino-Witten model and the non-Abelian T-dual of the principal chiral model. We present an auxiliary field deformation of the lambda model and exhibit a Lax connection whose flatness is equivalent to the remaining equations of motion after imposing the algebraic field equations. We also carry out a Hamiltonian analysis of the deformed theory, verifying that it obeys the assumptions of theorems in \cite{Bielli:2026ggs} which guarantee both (i) that the model enjoys the Maillet structure that ensures involution of its conserved charges, and (ii) that the theory possesses a classical Yangian. Thus, as in other cases, the auxiliary field deformation of the lambda model changes its classical dynamics while leaving the integrable structure essentially unchanged.
\end{abstract}

\vfill
\end{titlepage}
\hypersetup{pageanchor=true}

\tableofcontents
\bigskip
\hrule
\bigskip

\section{Introduction}
\label{sec:introduction}

Integrable field theories in two spacetime dimensions provide a setting in which strongly interacting dynamics can often be described exactly, both at the classical level and at the quantum level. Classical integrability is frequently studied using a Lax connection $\mathfrak{L}_\mu ( z )$ which is flat on-shell and which depends on an additional complex-valued spectral parameter $z \in \mathbb{C}$. The corresponding monodromy matrix,
\begin{align}
    \mathcal{M} ( z ) = \mathrm{Pexp} \left( - \int d \sigma \, \mathfrak{L}_\sigma ( z ) \right) \, ,
\end{align}
generates infinitely many conserved quantities, while nonlocal currents often organize these charges into a Yangian structure. In this work, we will focus on integrable $2d$ sigma models. In such cases, in the Hamiltonian formulation, the spatial Lax matrix typically has a non-ultralocal Poisson bracket of Maillet type \cite{MAILLET198654,Maillet:1985ec}, which controls the involution of monodromy invariants. These structures are closely related, but each illuminates a different aspect of the model: the Lax connection encodes the equations of motion, the Yangian organizes nonlocal symmetries, and the Maillet bracket describes the algebra of spectral observables. For reviews of these ideas and their role in integrable sigma models, see \cite{Hoare:2021dix,Loebbert:2016cdm}.

A particularly interesting example is the lambda model,\footnote{Throughout, ``lambda model'' means the so-called isotropic lambda model, whose coupling matrix is $\lambda_{ab}=\lambda\delta_{ab}$. This contrasts with anisotropic lambda deformations, in which different Lie-algebra directions may carry different couplings \cite{Sfetsos:2014lla}.} which was introduced as an integrable deformation of the principal chiral model (PCM) \cite{Sfetsos:2013wia}. The physical degree of freedom in this theory is a group-valued map $g:\Sigma\to G$, but it is convenient to express the Lagrangian for this model in a ``gauge field form'' which involves an additional non-dynamical field $a_\pm$ valued in the Lie algebra $\mathfrak{g}$ of $G$. We will refer to the equivalent presentation of this theory obtained by integrating out $a_\pm$, expressing all quantities only in terms of $g$ and its associated left- and right-invariant Maurer-Cartan forms, as the ``sigma model form'' of the theory. Using the gauge field $a_\pm$, the Lax connection for the lambda model can be written in terms of a flat and conserved current $L_\pm$ as
\begin{gather}
 L_\pm=\frac{2}{1+\lambda}a_\pm\,,
 \qquad
 \Lax^{(0)}_\pm(z)=\frac{L_\pm}{1\mp z}\,.
 \label{eq:intro-lambda-current}
\end{gather}
Here $\lambda$ is the deformation parameter; we decorate the Lax connection $\mathfrak{L}_\pm^{(0)}$ with a ${}^{(0)}$ to indicate that it is associated with the standard lambda model, to be contrasted with auxiliary field deformations of this theory that will be introduced shortly.

The simplicity of \eqref{eq:intro-lambda-current} is accompanied by a rich Hamiltonian description: the canonical Wess-Zumino-Witten (WZW) currents form two commuting Kac-Moody algebras, and the spatial Lax matrix is governed by the standard lambda-model twist function. The lambda model therefore supplies a natural seed whose Lagrangian and Hamiltonian integrable structures are both under good control.

The lambda action is a gauged WZW model coupled to a gauge-fixed PCM, with $a_\pm$ playing the role of the covariant PCM current. The deformation parameter therefore appears in an algebraic sector of the action in gauge field form, and the physical sigma model metric and flux emerge only after the gauge field has been eliminated. Modifying this sector can produce new interactions while leaving the explicitly group-dependent WZW couplings unchanged.

Auxiliary field sigma models \cite{Ferko:2024ali} provide precisely this kind of construction. In these theories one introduces Lie-algebra-valued fields $v_\pm$ without derivatives and lets them interact through a differentiable function of paired trace invariants. Eliminating $v_\pm$ produces nonlinear sigma model interactions without adding propagating degrees of freedom. The original construction for the PCM used an arbitrary function of a single invariant and was shown to contain deformations generated by functions of the stress tensor, including the $T\overline T$ \cite{Zamolodchikov:2004ce,Smirnov:2016lqw,Cavaglia:2016oda} and root-$T\overline T$ \cite{Ferko:2022cix,Babaei-Aghbolagh:2022leo} flows, while preserving the classical Lax and Maillet structures. Its higher-spin extension promoted the interaction function to $E(\nu_2,\ldots,\nu_N)$, thereby incorporating bilinears of higher-spin currents \cite{Bielli:2024ach}. In both constructions, invariance of $E$ under independent conjugations of $v_+$ and $v_-$ implies a commutator identity that is independent of its detailed form.

The framework has since been extended in several complementary directions. The auxiliary field theory and its Wess-Zumino extension were derived from four-dimensional Chern-Simons theory in \cite{Fukushima:2024nxm}. Abelian T-duality was shown to commute with the $T\overline T$ deformation, while the auxiliary field formulation established the compatibility of non-Abelian T-duality with arbitrary $T\overline T$-like flows \cite{Bielli:2024khq,Bielli:2024ach}. Auxiliary field interactions were also combined with Yang--Baxter and bi-Yang--Baxter deformations in \cite{Bielli:2024fnp}. Spin-two auxiliary deformations of general $\mathbb Z_N$ coset models and their Hamiltonian integrability were developed in \cite{Cesaro:2024ipq}. Higher-spin interactions were extended to $\mathbb Z_2$ symmetric and $\mathbb Z_4$ semi-symmetric spaces, including a Wess-Zumino coupling in the latter case, in \cite{Bielli:2024oif}. The same framework was subsequently applied to dimensionally reduced gravity \cite{Cesaro:2025msv}.

Recent work has also clarified the conserved currents, integrable flows, and alternative auxiliary formulations of these theories. The local higher-spin currents and their Smirnov-Zamolodchikov flows were studied systematically in \cite{Bielli:2025uiv}. Geometric interpretations were developed for auxiliary field deformations of the $\mathbb{CP}^{N-1}$ model \cite{Ferko:2025bhv}, and Lie-algebra-valued and scalar auxiliary formulations were related by a Legendre transformation in \cite{Baglioni:2025tsc}. The construction for dimensionally reduced gravity was subsequently reformulated in complementary duality frames and uplifted to four dimensions \cite{Bielli:2026vit}.

A common feature of auxiliary field deformations of sigma models is the appearance of two distinguished one-forms, one of which is flat and one of which is conserved. This generalizes the mechanism underlying integrability for many $2d$ sigma models, in which one can identify a single current which is both flat and conserved. In particular, the original construction of Brezin, Itzykson, Zinn-Justin, and Zuber (BIZZ) builds nonlocal currents from a single Lie-algebra-valued one-form that is flat and conserved \cite{BREZIN1979442}. For the auxiliary field models, although flatness and conservation are now distributed between two compatible one-forms, the generalized construction of \cite{Bielli:2026ggs} shows that this is still enough to generate an infinite conserved tower; this work also gives algebraic conditions under which the first two charges satisfy the classical Yangian relations. For the lambda model, one must also determine whether the auxiliary interactions preserve the Hamiltonian current algebra needed for Yangian symmetry and Maillet integrability.

It is then natural to ask whether auxiliary field interactions can be added to the lambda model while preserving its classical integrable structure. The question has three parts. First, can one construct a deformation whose equations admit a Lax representation for every interaction function $E(\nu_2,\ldots,\nu_N)$? Second, does the same family retain the Yangian and Maillet structures of the standard lambda model? Third, how does the construction behave under the limit from the lambda model to the non-Abelian T-dual PCM? The second question is particularly restrictive, since it depends not only on the equations of motion but also on the Dirac brackets obtained after the algebraic constraints have been eliminated.

The three questions probe different structures. The auxiliary interaction modifies the equations that relate $v_\pm$, $a_\pm$, and the group field, while the fundamental Kac-Moody brackets originate in the WZW part of the action in gauge field form. The task is therefore to express the current components entering the spatial Lax matrix as linear combinations of the canonical WZW currents and compute their brackets.

In this work we answer these questions affirmatively at the classical level. We construct auxiliary field deformations of the lambda model and determine their non-Abelian T-dual (NATD) limit. The Lax structure persists for the full family, and likewise the Yangian and Maillet structures generically survive under the assumption that certain algebraic constraints can be eliminated.

The structure of the paper is as follows. In Section~\ref{sec:review} we review the auxiliary field mechanism and the lambda model. In Section~\ref{sec:af-lambda} we construct the deformed action, derive its equations, give its leading current-bilinear interpretation, and derive its sigma model form and non-Abelian T-dual limit. Section~\ref{sec:classical-integrability} establishes the Lax representation and generalized BIZZ tower, derives the Hamiltonian current algebra, and applies it to the Yangian charges and the Maillet bracket. Section~\ref{sec:conclusion} summarizes our results and discusses several open questions. Appendices \ref{app:eoms-lax}, \ref{app:hamiltonian-constraints}, and \ref{app:yangian-maillet} collect certain longer calculations which are relevant for the Lagrangian, Hamiltonian, Yangian, and Maillet analyses.


\section{Review of auxiliary field deformations and lambda model}
\label{sec:review}

We begin by briefly reviewing some generalities about auxiliary field deformations and the standard lambda model, respectively, which will be combined in Section \ref{sec:af-lambda}. Our goal is to establish some basic properties which will be useful later, while also fixing the conventions used throughout the paper.

We work on a flat Lorentzian worldsheet $\Sigma$ with coordinates $(\tau,\sigma)$, and define
\begin{gather}
 x^\pm=\frac12(\tau\pm\sigma)\,,
 \qquad
 \partial_\pm=\partial_\tau\pm\partial_\sigma\,,
 \qquad
 X_\pm=X_\tau\pm X_\sigma \, ,
 \label{eq:worldsheet-conventions}
\end{gather}
where $X_\mu$ is a generic one-form. Our Hodge star convention is $\star\dd\tau=\dd\sigma$ and $\star\dd\sigma=\dd\tau$, so that $\star^2=1$ on one-forms. For Lie-algebra-valued forms the bracket includes the wedge product. We write $\doteq$ for equality after imposing only the algebraic equations for $v_\pm$, and $\approx$ to indicate equality after all relevant Euler-Lagrange equations have been used.

These conventions translate directly between components and forms. If $X=X_+\dd x^++X_-\dd x^-$, then $\dd(\star X)=0$ is the conservation equation $\partial_+X_-+\partial_-X_+=0$, whereas the $\dd x^+\wedge\dd x^-$ component of $\dd X+\frac12[X,X]$ is $\partial_+X_- -\partial_-X_+ +[X_+,X_-]$. We will primarily work at the component level using light-cone coordinates, although we sometimes give the corresponding formulas in form notation when it is helpful for clarity or brevity.

Let $G$ be a Lie group with semisimple Lie algebra $\mathfrak g$. We choose generators $T_A$ satisfying $[T_A,T_B]=f_{AB}{}^C T_C$, and write $\gamma_{AB}=\tr(T_AT_B)$ for a nondegenerate invariant bilinear form.

\subsection{Auxiliary field deformations}
\label{subsec:af-review}

Auxiliary field sigma models modify a seed theory by coupling to Lie-algebra-valued fields $v_\pm$ that enter without derivatives. For the higher-spin family of \cite{Bielli:2024ach}, the interaction function is a differentiable function of the paired invariants
\begin{gather}
 \nu_n=\tr(v_+^n) \tr(v_-^n)\,,
 \qquad
 E=E(\nu_2,\ldots,\nu_N)\,.
 \label{eq:af-invariants}
\end{gather}
In fact, one can also consider more general interaction functions $E = E \left( \tr ( v_\pm^n ) \right)$, including those depending on ``mixed-power'' combinations such as $\left( \tr ( v_+^2 ) \right)^3 \left( \tr ( v_-^3 ) \right)^2$, and later works typically consider this broader class of interaction functions \cite{Bielli:2025uiv,Bielli:2026ggs}. However, for simplicity, in the present article we will restrict attention to functions of the form (\ref{eq:af-invariants}).

The two factors in $\nu_n$ have opposite Lorentz weights, so their product is a scalar. For a semisimple algebra the list begins at $n=2$. We take traces in a fixed $d$-dimensional matrix representation and choose $N$ so that the traces through order $N$ generate all higher single-field traces $\tr(v_\pm^n)$. By the Cayley-Hamilton theorem, $N=d$ suffices, though a smaller value may also suffice.\footnote{However, these traces need not generate all invariants of $\mathfrak g$. For instance, for $\mathfrak{so}(2n)$, the Pfaffian is invariant under $SO(2n)$ conjugation but changes sign under conjugation by an orthogonal reflection.
Traces of all powers remain unchanged under that reflection.
Thus the traces determine the Pfaffian's square, $\det v$, but cannot determine its sign.} Each trace, and hence $E$, is unchanged under independent conjugations of $v_+$ and $v_-$. We require $E$ to be differentiable. This invariance implies the commutator identity used below.

The PCM provides a simple illustration. Let $j_\pm=g^{-1}\partial_\pm g$. In a common normalization its auxiliary field Lagrangian is
\begin{gather}
 \mathcal L_{\mathrm{AFSM}}
 =\frac12\tr(j_+j_-)+\tr(v_+v_-)
 +\tr(j_+v_-+j_-v_+)+E(\nu_2,\ldots,\nu_N)\,.
 \label{eq:afsm-review-action}
\end{gather}
Although eliminating $v_\pm$ generally produces a nonlinear action for $g$, the auxiliary field equations imply a simple identity that follows from conjugation invariance of $E$.

Let the derivatives of the interaction function with respect to the auxiliary fields be
\begin{gather}
 \Delta_\pm=\frac{\delta E}{\delta v_\mp}\,.
 \label{eq:delta-definition}
\end{gather}
Conjugating $v_\mp$ alone and using nondegeneracy of the bilinear form gives
\begin{gather}
 [\Delta_\pm,v_\mp]=0\,.
 \label{eq:delta-commutator}
\end{gather}
Indeed, varying $v_\mp$ by $[X,v_\mp]$ pairs $[v_\mp,\Delta_\pm]$ with an arbitrary $X\in\mathfrak g$.\footnote{This step uses only the fact that $E$ is unchanged when $v_\mp$ is conjugated while the other auxiliary field is held fixed. It does not assume that this conjugation is a symmetry of the full action.}

The labels in \eqref{eq:delta-definition} follow the light-cone pairing in the Lagrangian: varying $v_-$ produces the equation with a plus label, and conversely. Equation \eqref{eq:delta-commutator} states that $\Delta_\pm$ commutes with the field $v_\mp$ with respect to which $E$ was differentiated.

Varying \eqref{eq:afsm-review-action} with respect to $v_\mp$ gives
\begin{gather}
 j_\pm+v_\pm+\Delta_\pm\doteq0\,.
 \label{eq:afsm-aux-eom}
\end{gather}
Taking its commutator with $v_\mp$ and using \eqref{eq:delta-commutator}, we obtain
\begin{gather}
 [v_\mp,j_\pm]\doteq[v_\pm,v_\mp]\,.
 \label{eq:afsm-fundamental-identity}
\end{gather}
The complete auxiliary equation depends on $E$, whereas \eqref{eq:afsm-fundamental-identity} has the same form for every interaction function in \eqref{eq:af-invariants}. We note that this commutator identity can also be derived from the explicit functional form of $\Delta_\pm$ and the generalized Jacobi identity; see, for instance, Section 2.2 and Appendix B of \cite{Bielli:2024ach}.

The commutator identity, and its analogues in other auxiliary field sigma models, is a key ingredient in establishing classical integrability. In the PCM example, variation of $g$ gives
\begin{gather}
 \partial_+(j_-+2v_-)+\partial_-(j_++2v_+)
 \approx2\bigl([v_-,j_+]+[v_+,j_-]\bigr)\,.
 \label{eq:afsm-group-eom}
\end{gather}
The two commutators on the right cancel after \eqref{eq:afsm-fundamental-identity} is used. The group equation therefore becomes the conservation law
\begin{gather}
 \partial_+\mathfrak{J}_-+\partial_-\mathfrak{J}_+\approx0\,,
 \qquad
 \mathfrak{J}_\pm=-(j_\pm+2v_\pm)\,.
 \label{eq:afsm-conserved-current}
\end{gather}
At the same time, $j=g^{-1}\dd g$ is flat by the Maurer-Cartan identity, and the auxiliary equation relates the commutators of $j$ and $\mathfrak{J}$. As a result, after imposing the auxiliary field equations, the group equation is equivalent to flatness of the Lax connection
\begin{align}
    \mathfrak{L}^{\text{(PCM)}}_\pm = \frac{j_\pm \pm z \mathfrak{J}_\pm}{1 - z^2} \, .
\end{align}
As we will see, the same pattern occurs in the lambda model: the interaction function changes the relation between the physical fields and the currents, while adjoint invariance supplies the algebraic identity needed for integrability. More explicitly, the same use of \eqref{eq:afsm-fundamental-identity} shows that the commutator of two $\mathfrak{J}$'s equals that of two Maurer-Cartan currents, together with the corresponding mixed compatibility relation. Flatness of $j$, conservation of $\mathfrak{J}$, and these two algebraic identities are the four inputs of the generalized BIZZ construction \cite{Bielli:2026ggs}. We will recover their lambda model analogues directly in Section~\ref{subsec:eom-currents}.

This review also explains why we will keep the auxiliary fields throughout the derivation. Eliminating them at the outset would replace the transparent identity \eqref{eq:afsm-fundamental-identity} by nonlinear relations among the physical currents. The auxiliary field formulation exposes the integrable mechanism before any model-dependent interaction function is chosen.

\subsection{The lambda model}
\label{subsec:lambda-review}

We next review the gauge field form of the lambda model in the conventions of \cite{Sfetsos:2013wia}. Introduce the left- and right-invariant currents
\begin{gather}
 j_\pm=g^{-1}\partial_\pm g\,,
 \qquad
 k_\pm=-(\partial_\pm g)g^{-1}\,,
 \qquad
 \Ad_gX=gXg^{-1}\,.
 \label{eq:lambda-maurer-cartan}
\end{gather}
In terms of a nondynamical gauge field $a_\pm$, the action is
\begin{align}
 S_\lambda(g,a)
 &=-\frac{\kappa}{2}\int_\Sigma d^2x\,\tr(j_+j_-)
 +\frac{\kappa}{6}\int_{M_3}d^3x\,
   \epsilon^{ijk}\tr\bigl(j_i[j_j,j_k]\bigr)
 \notag\\
 &\quad+\kappa\int_\Sigma d^2x\,\tr\bigl(
 a_+j_-+a_-k_+ +a_+g^{-1}a_-g-\lambda^{-1}a_+a_-
 \bigr) \, .
 \label{eq:lambda-action}
\end{align}
Here $\partial M_3=\Sigma$ and $\kappa$ is the WZW coefficient. Our normalization is related to the integer level, when the usual quantization conditions apply, by $k_{\mathrm{WZW}}=4\pi\kappa$ \cite{Witten:1983tw}.

The second integral is the Wess-Zumino term \cite{Wess:1971yu}, written using an extension of $g$ to a three-manifold $M_3$ with boundary $\Sigma$. Only its standard local variation is needed below. Notice also that $a_\pm$ appears without derivatives. Its equations are algebraic in the Lagrangian description, even though the associated constraints are essential to the Hamiltonian current algebra.

The form of \eqref{eq:lambda-action} follows by coupling a gauged WZW model to a PCM whose left action has been gauged \cite{Sfetsos:2013wia}. After fixing the PCM field to the identity, its covariant Maurer-Cartan current becomes $a_\pm$, and the action decomposes as
\begin{gather}
 S_\lambda(g,a)
 =S_{\mathrm{gWZW}}(g,a)+S_{\mathrm{gPCM}}^{\mathrm{gf}}(a)\,,
 \qquad
 S_{\mathrm{gPCM}}^{\mathrm{gf}}(a)
 =-\frac h2\int_\Sigma d^2x\,\tr(a_+a_-)\,,
 \label{eq:lambda-splitting}
\end{gather}
where
\begin{gather}
 1+\frac{h}{2\kappa}=\lambda^{-1}\,.
 \label{eq:lambda-pcm-sector}
\end{gather}
Before gauge fixing, the PCM contribution contains a second group-valued field $\widehat g$ through the covariant current $\widehat g^{-1}(\partial_\pm+a_\pm)\widehat g$. The local left action may be used to set $\widehat g=1$, after which this current becomes $a_\pm$ and produces the second term in \eqref{eq:lambda-splitting}. This derivation identifies the algebraic $a_+a_-$ coupling as the gauge-fixed remnant of the PCM kinetic term. Thus the PCM sector changes only the coefficient of $\tr(a_+a_-)$. We will introduce the auxiliary fields into this algebraic sector and leave all explicitly $g$-dependent gauged-WZW couplings unchanged.

The equations of the undeformed lambda model therefore imply that the combination
\begin{gather}
 L_\pm=\frac{2}{1+\lambda}a_\pm\,,
 \label{eq:lambda-seed-current}
\end{gather}
is both flat and conserved on-shell, i.e.
\begin{gather}
 \partial_+L_- -\partial_-L_+ +[L_+,L_-]\approx0\,,
 \qquad
 \partial_+L_-+\partial_-L_+\approx0\,.
 \label{eq:lambda-seed-identities}
\end{gather}
These two equations are equivalent to the flatness of the Lax connection \cite{Sfetsos:2013wia}
\begin{gather}
 \Lax^{(0)}_\pm(z)=\frac{L_\pm}{1\mp z}\,.
 \label{eq:lambda-seed-lax}
\end{gather}
To see this directly, note that the curvature of \eqref{eq:lambda-seed-lax} is
\begin{align}
 \mathcal F^{(0)}_{+-}(z)
 &=
 \partial_+\Lax^{(0)}_-(z)
 -\partial_-\Lax^{(0)}_+(z)
 +[\Lax^{(0)}_+(z),\Lax^{(0)}_-(z)]
 \notag\\
 &=
 \frac{1}{1-z^2}
 \left[
 \partial_+L_- -\partial_-L_+ +[L_+,L_-]
 -z\bigl(\partial_+L_-+\partial_-L_+\bigr)
 \right]\,.
\end{align}
The constant and linear terms in the numerator reproduce, respectively, the flatness and conservation equations in \eqref{eq:lambda-seed-identities}. Hence the connection is flat for every value of the spectral parameter precisely when both equations are satisfied. This familiar calculation anticipates the deformed construction, where the same two powers of $z$ will instead be controlled by two compatible currents.

In the auxiliary field deformation, the single flat and conserved current $L_\pm$ is replaced by two one-forms: one is flat but need not be conserved, and the other is conserved but need not be flat. Their coincidence with $L$ when $E=0$ will give a direct check of both the current normalization and the Lax representation.

The two ingredients of this section are now in a compatible form. The gauge-fixed PCM sector identifies $a_\pm$ as the current to which new algebraic fields should couple, while the invariant auxiliary construction shows which combinations of $v_\pm$ preserve the commutator identities. We will use these observations to define the deformation without modifying the gauged-WZW part of the action.

\section{Auxiliary field deformation of the lambda model}
\label{sec:af-lambda}

We now combine the two ingredients reviewed in Section~\ref{sec:review}. The new terms are confined to the gauge-fixed PCM sector. At $E=0$, eliminating $v_\pm$ recovers \eqref{eq:lambda-action}. We will then show that the deformed equations retain the current identities needed for classical integrability.

The construction is most transparent in the variables $(g,a_\pm,v_\pm)$. The group field carries the propagating degrees of freedom, while the gauge and auxiliary fields are determined algebraically. We use a commutator identity implied by the auxiliary field equations, which holds for the entire class of models \eqref{eq:af-invariants}.

\subsection{Definition and standard lambda limit}
\label{subsec:definition}

It is convenient to define
\begin{gather}
 q_\lambda=\lambda-\lambda^{-1}\,.
 \label{eq:q-lambda}
\end{gather}
Equivalently, $q_\lambda=-(1-\lambda^2)/\lambda$. We work with $0<\lambda<1$ and $\kappa\neq0$, so $q_\lambda\neq0$ and the linear transformations used below are invertible. The coefficient $q_\lambda$ measures the difference between the quadratic gauge field coupling used in the new action in gauge field form and the coupling of the ordinary lambda model.

The auxiliary field deformation of the lambda model is
\begin{align}
 S_{\mathrm{AF}\text{-}\lambda}
 &=-\frac{\kappa}{2}\int_\Sigma d^2x\,\tr(j_+j_-)
 +\frac{\kappa}{6}\int_{M_3}d^3x\,
   \epsilon^{ijk}\tr\bigl(j_i[j_j,j_k]\bigr)
 \notag\\
 &\quad+\kappa\int_\Sigma d^2x\,\tr\bigl(
 a_+j_-+a_-k_+ +a_+g^{-1}a_-g-\lambda a_+a_-
 \bigr)
 \notag\\
 &\quad+\kappa q_\lambda\int_\Sigma d^2x\,
 \left[
 \tr\bigl(a_+v_-+a_-v_+-v_+v_-\bigr)
 -E(\nu_2,\ldots,\nu_N)
 \right]\,.
 \label{eq:af-lambda-action}
\end{align}
The WZW terms and the three explicitly $g$-dependent gauge couplings are the same as in \eqref{eq:lambda-action}. The compensating coefficient of $a_+a_-$ is now $-\lambda$, while $v_\pm$ enter only algebraically.

Equation \eqref{eq:af-lambda-action} replaces the gauge-fixed PCM term by an auxiliary field presentation while leaving the gauged WZW action unchanged. The common factor $\kappa q_\lambda$ drops out of the auxiliary field equations \eqref{eq:af-lambda-aux-eom}. Since neither $a_\pm$ nor $v_\pm$ carries a derivative, the new action introduces no additional propagating fields.

Variation with respect to $v_\mp$ gives
\begin{gather}
 a_\pm\doteq v_\pm+\Delta_\pm\,,
 \qquad
 \Delta_\pm=\frac{\delta E}{\delta v_\mp}\,.
 \label{eq:af-lambda-aux-eom}
\end{gather}
When $E=0$, this equation sets $v_\pm\doteq a_\pm$. The last line of \eqref{eq:af-lambda-action} then contributes $q_\lambda\tr(a_+a_-)$, and the coefficients of the terms multiplying $a_+ a_-$ combine as $-\lambda+q_\lambda=-\lambda^{-1}$, so we recover
\begin{gather}
 \left.S_{\mathrm{AF}\text{-}\lambda}\right|_{E=0}
 \doteq S_\lambda\,.
 \label{eq:af-lambda-e0}
\end{gather}
This check fixes the coefficient $-\lambda$ in the second line of \eqref{eq:af-lambda-action}. The minus sign multiplying $E$ gives the auxiliary equation $a_\pm\doteq v_\pm+\Delta_\pm$ with our definition of $\Delta_\pm$. Taking the commutator of the plus equation with $v_-$ and of the minus equation with $v_+$, and using \eqref{eq:delta-commutator}, gives
\begin{gather}
 [v_\pm,a_\mp]\doteq[v_\pm,v_\mp]\,.
 \label{eq:af-lambda-fundamental-commutator}
\end{gather}
All dependence on the derivatives of $E$ has disappeared. This identity simplifies both the differential and algebraic relations obeyed by the currents.

\subsection{Equations of motion and currents}
\label{subsec:eom-currents}

Because the deformation changes no term with explicit $g$ dependence, the group equation has the same form as in the ordinary lambda model \cite{Sfetsos:2013wia}. Its two equivalent presentations are
\begin{align}
 &\partial_+\bigl(g^{-1}\partial_-g+g^{-1}a_-g\bigr)
 -\partial_-a_+
 +\bigl[a_+,g^{-1}\partial_-g+g^{-1}a_-g\bigr]=0\,,
 \notag\\
 &\partial_-\bigl(-\partial_+gg^{-1}+ga_+g^{-1}\bigr)
 -\partial_+a_-
 +\bigl[a_-,-\partial_+gg^{-1}+ga_+g^{-1}\bigr]=0\,.
 \label{eq:af-lambda-g-eom}
\end{align}
The gauge field equations, on the other hand, contain the new auxiliary sources:
\begin{align}
 -(\lambda-\Ad_g)a_+-\partial_+gg^{-1}+q_\lambda v_+&=0\,,
 \notag\\
 -(\lambda-\Ad_g^{-1})a_-+g^{-1}\partial_-g+q_\lambda v_-&=0\,.
 \label{eq:af-lambda-gauge-eom}
\end{align}
The variations leading to these expressions are collected in Appendix~\ref{app:eoms-lax}.

The two group equations are related by conjugation with $g$, up to an overall sign, and express the same Euler--Lagrange equation in left- and right-invariant variables. Both forms are useful because the gauge field equations express the corresponding Maurer-Cartan currents in terms of $a_\pm$ and $v_\pm$. Explicitly, \eqref{eq:af-lambda-gauge-eom} gives
\begin{gather}
 -\partial_+gg^{-1}=(\lambda-\Ad_g)a_+-q_\lambda v_+\,,
 \qquad
 g^{-1}\partial_-g=(\lambda-\Ad_g^{-1})a_--q_\lambda v_-\,.
 \label{eq:af-lambda-g-solutions}
\end{gather}
This step removes $g$ from the remaining differential calculation; its dependence is already encoded in the algebraic gauge constraints.

Substituting these expressions into \eqref{eq:af-lambda-g-eom}, the two forms of the group equation become
\begin{align}
 \partial_+(\lambda a_--q_\lambda v_-)-\partial_-a_+
 +[a_+,\lambda a_--q_\lambda v_-]&=0\,,
 \notag\\
 \partial_-(\lambda a_+-q_\lambda v_+)-\partial_+a_-
 +[a_-,\lambda a_+-q_\lambda v_+]&=0\,.
 \label{eq:af-lambda-physical-eoms}
\end{align}
The two lines of \eqref{eq:af-lambda-physical-eoms} are especially convenient because every commutator is built from $a_\pm$ and $v_\pm$. Adding them, the terms proportional to $[a_+,a_-]$ cancel directly, while the mixed terms cancel by \eqref{eq:af-lambda-fundamental-commutator}. Subtracting them instead doubles the curvature-like terms. Thus the same algebraic identity that removed $\Delta_\pm$ now separates conservation from flatness.

Their sum and difference reveal the two current equations. After applying \eqref{eq:af-lambda-fundamental-commutator}, one finds
\begin{gather}
 (\lambda-1)(\partial_+a_-+\partial_-a_+)
 -q_\lambda(\partial_+v_-+\partial_-v_+)\approx0\,,
 \label{eq:af-lambda-sum}
\end{gather}
and
\begin{align}
 &(1+\lambda)(\partial_+a_--\partial_-a_+)
 -q_\lambda(\partial_+v_--\partial_-v_+)
 +2\lambda[a_+,a_-]-2q_\lambda[v_+,v_-]\approx0\,.
 \label{eq:af-lambda-difference}
\end{align}
The sum becomes a conservation law, while the difference has the nonlinear terms required for a curvature. Appendix~\ref{app:eoms-lax} provides further details on the derivation of these equations.

Motivated by these two combinations, define
\begin{align}
 \mathcal A_\pm
 &=\frac{2}{1+\lambda}
   \bigl(\lambda a_\pm+(1-\lambda)v_\pm\bigr)\,,
 \notag\\
 \mathcal B_\pm
 &=\frac{2}{1+\lambda}
   \bigl((1+\lambda)v_\pm-\lambda a_\pm\bigr)\,.
 \label{eq:split-currents}
\end{align}
The coefficients are fixed by three simple requirements. The sum equation must become conservation of $\mathcal B$, the difference equation must become flatness of $\mathcal A$, and both currents must reduce to $2a_\pm/(1+\lambda)$ when $v_\pm=a_\pm$. These requirements determine the linear combinations in \eqref{eq:split-currents}, up to a common convention already fixed by \eqref{eq:lambda-seed-current}.

This linear transformation is invertible for the parameter range used below, with inverse
\begin{gather}
 v_\pm=\frac{1+\lambda}{4}(\mathcal A_\pm+\mathcal B_\pm)\,,
 \qquad
 a_\pm=\frac{1+\lambda}{4\lambda}
 \bigl((1+\lambda)\mathcal A_\pm
 -(1-\lambda)\mathcal B_\pm\bigr)\,.
 \label{eq:split-current-inverse}
\end{gather}
At $E=0$, where $v_\pm=a_\pm$, both currents reduce to the standard lambda current: $\mathcal A_\pm=\mathcal B_\pm=L_\pm$. The inverse relation \eqref{eq:split-current-inverse} will be equally useful in the Hamiltonian analysis. It allows every algebraic constraint written in terms of $a_\pm$ and $v_\pm$ to be expressed using $\mathcal A$ and $\mathcal B$, without solving the nonlinear auxiliary equation for either field.

Using $q_\lambda=(\lambda-1)(\lambda+1)/\lambda$, the sum equation \eqref{eq:af-lambda-sum} is equivalent to
\begin{gather}
 \partial_+\mathcal B_-+\partial_-\mathcal B_+\approx0\,.
 \label{eq:split-conservation}
\end{gather}
The difference equation gives the curvature of $\mathcal A$. The only nontrivial algebraic step is
\begin{gather}
 [\mathcal A_+,\mathcal A_-]\doteq
 \frac{4}{(1+\lambda)^2}
 \left(\lambda^2[a_+,a_-]+(1-\lambda^2)[v_+,v_-]\right)\,,
 \label{eq:split-a-commutator}
\end{gather}
which follows from \eqref{eq:af-lambda-fundamental-commutator}. Substitution into \eqref{eq:af-lambda-difference} gives
\begin{gather}
 \partial_+\mathcal A_- -\partial_-\mathcal A_+
 +[\mathcal A_+,\mathcal A_-]\approx0\,.
 \label{eq:split-flatness}
\end{gather}
Thus $\mathcal A$ is flat and $\mathcal B$ is conserved.

The two currents also obey algebraic compatibility relations. Expanding their commutators and using only \eqref{eq:af-lambda-fundamental-commutator}, we obtain
\begin{align}
 [\mathcal A_+,\mathcal A_-]
 &\doteq[\mathcal B_+,\mathcal B_-]\,,
 \notag\\
 [\mathcal A_+,\mathcal B_-]
 &\doteq[\mathcal B_+,\mathcal A_-]\,.
 \label{eq:split-component-commutators}
\end{align}
These identities are independent of the detailed form of $E$. The auxiliary interaction changes the algebraic relation $a_\pm\doteq v_\pm+\Delta_\pm$, while the linear transformation \eqref{eq:split-currents} from $(a,v)$ to $(\mathcal A,\mathcal B)$ is independent of $E$.

The first identity equates the quadratic terms that will appear in the Lax curvature. The second removes the term linear in the spectral parameter that would otherwise mix $\mathcal A$ and $\mathcal B$. Together with \eqref{eq:split-conservation} and \eqref{eq:split-flatness}, these identities establish the Lax representation without specifying the detailed form of $E$.

\subsection{Equivalent presentation and perturbative interpretation}
\label{subsec:shifted}

The action \eqref{eq:af-lambda-action} is well adapted to the current calculation, but a simple field redefinition makes its relation to the ordinary lambda model more transparent. Set
\begin{gather}
 w_\pm=v_\pm-a_\pm\,.
 \label{eq:shifted-auxiliary}
\end{gather}
Cyclicity of the trace gives
\begin{gather}
 \tr(a_+v_-+a_-v_+-v_+v_-)
 =\tr(a_+a_--w_+w_-)\,.
 \label{eq:af-lambda-shifted-trace}
\end{gather}
The first term on the right combines with $-\lambda a_+a_-$, and the full action becomes
\begin{align}
 S_{\mathrm{AF}\text{-}\lambda}
 =S_\lambda(g,a)
 -\kappa q_\lambda\int_\Sigma d^2x\,
 \left[
 \tr(w_+w_-)
 +E\bigl(\nu_2(a+w),\ldots,\nu_N(a+w)\bigr)
 \right]\,,
 \label{eq:af-lambda-shifted-action}
\end{align}
where $\nu_n(a+w)=\tr((a_++w_+)^n)\tr((a_-+w_-)^n)$. This expression is exactly equivalent to \eqref{eq:af-lambda-action}; no equation of motion has been used. It isolates the standard lambda action and places the deformation in a purely algebraic sector.

In these variables the auxiliary equations take the compact form
\begin{gather}
 w_\pm+\Delta_\pm[a+w]\doteq0\,,
 \label{eq:shifted-aux-eom}
\end{gather}
where the notation indicates that $\Delta_\pm$ is evaluated at $v_\pm=a_\pm+w_\pm$. In particular, $w_\pm=0$ when $E=0$. We retain the unshifted variables for the current calculation because their commutator identity is simpler, but \eqref{eq:af-lambda-shifted-action} is better suited to an expansion around the undeformed lambda model.

The shifted variables also reveal the leading physical perturbation. Choose a fixed $n\geq2$ and take
\begin{gather}
 E=\varepsilon\nu_n\,.
 \label{eq:epsilon-potential}
\end{gather}
The shifted auxiliary equation gives $w_\pm=O(\varepsilon)$, and hence $\tr(w_+w_-)=O(\varepsilon^2)$. To first order, the interaction function in \eqref{eq:af-lambda-shifted-action} may therefore be evaluated at $v_\pm=a_\pm$. Using $a_\pm=(1+\lambda)L_\pm/2$, we find
\begin{align}
 S_{\mathrm{AF}\text{-}\lambda}
 \doteq S_\lambda
 &-\kappa q_\lambda\varepsilon
 \left(\frac{1+\lambda}{2}\right)^{2n}
 \int_\Sigma d^2x\,
 \tr(L_+^n)\tr(L_-^n) +O(\varepsilon^2) \,.
 \label{eq:leading-current-bilinear}
\end{align}
There are two contributions from the shifted fields that might have entered at this order. The explicit quadratic term begins at $O(\varepsilon^2)$, while replacing $a_\pm$ by $a_\pm+w_\pm$ inside $E$ produces a correction of order $\varepsilon w=O(\varepsilon^2)$. Consequently the coefficient displayed in \eqref{eq:leading-current-bilinear} is obtained simply by evaluating the interaction function on the undeformed algebraic solution for the auxiliary and gauge fields.

The higher-spin currents $\tr(L_+^n)$ and $\tr(L_-^n)$ in this formula are chiral. Indeed, \eqref{eq:lambda-seed-identities} and cyclicity of the trace imply
\begin{gather}
 \partial_-\tr(L_+^n)
 =\frac n2\tr\bigl(L_+^{n-1}[L_+,L_-]\bigr)=0\,,
 \qquad
 \partial_+\tr(L_-^n)
 =-\frac n2\tr\bigl(L_-^{n-1}[L_+,L_-]\bigr)=0\,.
 \label{eq:lambda-higher-spin-chiral}
\end{gather}
Here we used $\partial_-L_+=\frac12[L_+,L_-]$ and $\partial_+L_-=-\frac12[L_+,L_-]$, which follow by adding and subtracting the two equations in \eqref{eq:lambda-seed-identities}.

Equation \eqref{eq:leading-current-bilinear} therefore shows that, to leading order in the deformation parameter, an auxiliary field deformation with a monomial interaction function represents an infinitesimal Smirnov-Zamolodchikov flow \cite{Smirnov:2016lqw} driven by a bilinear in higher-spin currents. In particular, the higher-spin currents triggering the flow are built from powers of the flat and conserved currents in the seed model. A similar structure has been observed in other examples of auxiliary field deformations; see, e.g., \cite{Bielli:2024ach,Bielli:2024fnp}.

This perturbative interpretation complements the exact current construction. The shifted action (\ref{eq:af-lambda-shifted-action}) is convenient for identifying the operator that initiates an infinitesimal deformation, whereas we have used the unshifted equations to establish the current identities for an arbitrary function $E$, without expanding in its couplings.

\subsection{Sigma model form and non-Abelian T-dual limit}
\label{subsec:sigma-natd}

We now integrate out the gauge fields, obtaining a sigma model action for the group field and a direct route to the non-Abelian T-dual limit.  Define
\begin{gather}
 \Omega_-=(\lambda-\Ad_g^{-1})^{-1}\,,
 \qquad
 \Omega_+=(\lambda-\Ad_g)^{-1}\,.
 \label{eq:af-lambda-omega}
\end{gather}
All inverses in this subsection are understood wherever the corresponding algebraic operators are nonsingular.  Invariance of the trace gives $(\Ad_g)^T=\Ad_g^{-1}$, and hence $\Omega_-^T=\Omega_+$.  The two equations in \eqref{eq:af-lambda-gauge-eom} therefore have the solution
\begin{gather}
 a_-=\Omega_-\bigl(j_-+q_\lambda v_-\bigr)\,,
 \qquad
 a_+=\Omega_+\bigl(k_++q_\lambda v_+\bigr)\,.
 \label{eq:af-lambda-gauge-solutions}
\end{gather}
Substitution into \eqref{eq:af-lambda-action} gives
\begin{align}
 S_{\mathrm{AF}\text{-}\lambda}[g,v]
 ={}&S_{\mathrm{WZW}}[g]
 +\kappa\int_\Sigma d^2x\,
 \tr\!\left[
 \bigl(k_++q_\lambda v_+\bigr)
 \frac{1}{\lambda-\Ad_g^{-1}}
 \bigl(j_-+q_\lambda v_-\bigr)
 \right]
 \notag\\
 &-\kappa q_\lambda\int_\Sigma d^2x\,
 \left[\tr(v_+v_-)+E(\nu_2,\ldots,\nu_N)\right]\,.
 \label{eq:af-lambda-sigma-form}
\end{align}
Here $S_{\mathrm{WZW}}[g]$ denotes the WZW terms in \eqref{eq:af-lambda-action}.  Equation \eqref{eq:af-lambda-sigma-form} is the sigma model form of the deformed theory: only $g$ carries derivatives, while $v_\pm$ remain algebraic.  Varying these fields reproduces
\begin{gather}
 \Omega_-\bigl(j_-+q_\lambda v_-\bigr)=v_-+\Delta_-\,,
 \qquad
 \Omega_+\bigl(k_++q_\lambda v_+\bigr)=v_++\Delta_+\,,
 \label{eq:af-lambda-sigma-auxiliary-equations}
\end{gather}
which is the auxiliary equation \eqref{eq:af-lambda-aux-eom} after using \eqref{eq:af-lambda-gauge-solutions}.

Setting $E=0$ provides a direct check.  Equation \eqref{eq:af-lambda-sigma-auxiliary-equations} gives
\begin{gather}
 v_-=(\lambda^{-1}-\Ad_g^{-1})^{-1}j_-\,,
 \qquad
 v_+=(\lambda^{-1}-\Ad_g)^{-1}k_+\,.
 \label{eq:af-lambda-sigma-standard-solution}
\end{gather}
Substituting this result into \eqref{eq:af-lambda-sigma-form}, we recover
\begin{align}
 \left.S_{\mathrm{AF}\text{-}\lambda}\right|_{E=0}
 ={}&S_{\mathrm{WZW}}[g]
 -\kappa\int_\Sigma d^2x\,
 \tr\!\left(
 j_+\frac{\Ad_g^{-1}}{\lambda^{-1}-\Ad_g^{-1}}j_-
 \right)
 \notag\\
 ={}&-\frac{\kappa}{2}\int_\Sigma d^2x\,
 \tr\!\left(
 j_+\frac{\id+\lambda\Ad_g^{-1}}
 {\id-\lambda\Ad_g^{-1}}j_-
 \right) +\frac{\kappa}{6}\int_{M_3}d^3x\,
 \epsilon^{ijk}\tr\bigl(j_i[j_j,j_k]\bigr)\,.
 \label{eq:af-lambda-standard-pure-action}
\end{align}
This is the usual pure sigma model action for the lambda model \cite{Sfetsos:2013wia} in the Wess-Zumino convention of \eqref{eq:lambda-action}.

We can now take the non-Abelian T-dual limit.\footnote{See Section 6.1 of \cite{Hoare:2021dix} for a pedagogical review of the analogous limit for the undeformed lambda model.} Introduce
\begin{gather}
 \epsilon=\frac{h}{2\kappa}\,,
 \qquad
 \lambda=(1+\epsilon)^{-1}\,,
 \qquad
 g=e^{\epsilon X}\,,
 \qquad
 v_\pm=-V_\pm\,,
 \qquad
 \kappa\longrightarrow\infty\,,
 \label{eq:af-lambda-natd-scaling}
\end{gather}
with $h,X,a_\pm,V_\pm$ and the couplings in $E$ held fixed. This choice of $\lambda$ follows from \eqref{eq:lambda-pcm-sector}. It agrees with the exponential parametrization $\lambda=e^{-\epsilon}$ at the required order, since both give $\lambda=1-\epsilon+O(\epsilon^2)$. The sign in $g=e^{\epsilon X}$ is a convention: $X\to-X$ exchanges the two sign choices. The leading expansions needed in the limit are
\begin{align}
 \kappa q_\lambda \longrightarrow-h\,, &\qquad 
 j_\pm=\epsilon\partial_\pm X+O(\epsilon^2)\,,
 \notag\\
 k_\pm=-\epsilon\partial_\pm X+O(\epsilon^2)\,, &\qquad 
 \lambda-\Ad_g^{-1}
 =-\epsilon(\id-\ad_X)+O(\epsilon^2)\,.
 \label{eq:af-lambda-natd-expansion}
\end{align}
The relative sign between $v_\pm$ and $V_\pm$ cancels in every invariant $\nu_n$ constructed from either set of auxiliary fields, so we do not distinguish between $\nu_n ( v_\pm )$ and $\nu_n ( V_\pm )$. The WZW kinetic and Wess-Zumino terms vanish, while the remaining terms in \eqref{eq:af-lambda-action} become
\begin{gather}
 \begin{aligned}
 S_{\mathrm{AF}\text{-}\mathrm{NATD}}^{\mathrm{master}}
 =h\int_\Sigma d^2x\,
 \bigg\{&\frac12\tr\bigl(XF_{+-}(a)\bigr)
 +\tr\!\left(
 \frac12a_+a_-+V_+V_-+a_+V_-+a_-V_+
 \right)
 \\[-0.2em]
 &+E(\nu_2,\ldots,\nu_N)\bigg\}\,,
 \end{aligned}
 \label{eq:af-lambda-natd-master}
\end{gather}
where $F_{+-}(a)=\partial_+a_--\partial_-a_++[a_+,a_-]$.  This is the gauge-fixed master action for the auxiliary field-deformed non-Abelian T-dual PCM \cite{Bielli:2024ach,Bielli:2024khq}. Its gauge field equations are
\begin{gather}
 (\id\pm\ad_X)a_\pm
 =\pm\bigl(\partial_\pm X\mp2V_\pm\bigr)\,,
 \label{eq:af-lambda-natd-gauge-equations}
\end{gather}
and substitution gives
\begin{align}
 S_{\mathrm{AF}\text{-}\mathrm{NATD}}
 =h\int_\Sigma d^2x\,
 \bigg\{&\frac12\tr\!\left[
 (\partial_+X-2V_+)
 \frac{1}{\id-\ad_X}
 (\partial_-X+2V_-)
 \right] +\tr(V_+V_-)+E(\nu_2,\ldots,\nu_N)\bigg\}\,.
 \label{eq:af-lambda-natd-action}
\end{align}
Equation \eqref{eq:af-lambda-natd-action} is precisely the auxiliary field-deformed non-Abelian T-dual of the principal chiral model \cite{Bielli:2024khq,Bielli:2025uiv}. The current variables have the finite limits $\mathcal A_\pm\to a_\pm$ and $\mathcal B_\pm\to-(a_\pm+2V_\pm)$, so the Lax connection reduces to the one of the T-dual theory.  The auxiliary field construction and the lambda-to-NATD limit therefore commute.

Let us close this section by collecting the current identities that we have established. In differential form notation, \eqref{eq:split-flatness} and \eqref{eq:split-conservation} read
\begin{align}
 \dd\mathcal A+\frac12[\mathcal A,\mathcal A] \approx0 \, , \qquad 
 \dd(\star\mathcal B) \approx0\,,
 \label{eq:split-differential-identities}
\end{align}
while \eqref{eq:split-component-commutators} becomes
\begin{align}
 [\mathcal A,\mathcal A] \doteq[\mathcal B,\mathcal B]\,,
 \qquad
 [\mathcal A,\star\mathcal B]
 \doteq[\mathcal B,\star\mathcal A]\doteq0\,.
 \label{eq:split-commutator-identities}
\end{align}
The first line states that $\mathcal A$ is flat and $\mathcal B$ is conserved; the second supplies their algebraic compatibility. These four relations are exactly the input needed in Subsection~\ref{subsec:lax-integrability} to construct the Lax connection and the generalized BIZZ currents.

\section{Classical integrability}
\label{sec:classical-integrability}

We now consider the integrable structure of the auxiliary field deformed lambda model (\ref{eq:af-lambda-action}). The analysis proceeds in four stages. Subsection~\ref{subsec:lax-integrability} constructs the Lax connection and the generalized BIZZ tower, Subsection~\ref{subsec:current-algebra} derives the Hamiltonian current algebra, Subsection~\ref{subsec:yangian} establishes the classical Yangian symmetry, and Subsection~\ref{subsec:maillet} obtains the Maillet algebra and twist function.

\subsection{Lax integrability}
\label{subsec:lax-integrability}

We now use the two currents constructed in Section~\ref{sec:af-lambda} to establish classical integrability.  Recall that, after imposing the equations of motion, $\mathcal A$ is flat and $\mathcal B$ is conserved, while the auxiliary equations imply the two commutator identities in \eqref{eq:split-commutator-identities}.  These four relations suggest the Lax connection
\begin{gather}
 \Lax(z)=\frac{\mathcal A+z\star\mathcal B}{1-z^2}\,,
 \qquad
 \Lax_\pm(z)=\frac{\mathcal A_\pm\pm z\mathcal B_\pm}{1-z^2}\,.
 \label{eq:af-lambda-lax}
\end{gather}
Let us verify its flatness directly.  The quadratic term in the curvature is proportional to
\begin{gather}
 [\mathcal A+z\star\mathcal B,
   \mathcal A+z\star\mathcal B]
 = [\mathcal A,\mathcal A]
 +2z[\mathcal A,\star\mathcal B]
 -z^2[\mathcal B,\mathcal B]\,.
 \label{eq:af-lambda-lax-quadratic}
\end{gather}
The mixed term vanishes, whereas $[\mathcal B,\mathcal B]\doteq[\mathcal A,\mathcal A]$ converts the first and last terms into $(1-z^2)[\mathcal A,\mathcal A]$.  Including the exterior derivative, we therefore find
\begin{align}
 \dd\Lax+\frac12[\Lax,\Lax]
 &\doteq \frac{1}{1-z^2}
 \left(
 \dd\mathcal A+\frac12[\mathcal A,\mathcal A]
 +z\,\dd(\star\mathcal B)
 \right)
 \approx0\,.
 \label{eq:af-lambda-lax-curvature}
\end{align}
Thus the equations of motion admit a Lax representation for every differentiable function $E(\nu_2,\ldots,\nu_N)$. Conversely, the constant and linear terms in the numerator of \eqref{eq:af-lambda-lax-curvature} reproduce flatness of $\mathcal A$ and conservation of $\mathcal B$. Appendix~\ref{app:eoms-lax} collects the variations and current identities leading to this formula.

When $E=0$, the auxiliary equations give $\mathcal A_\pm\doteq\mathcal B_\pm\doteq L_\pm$, and hence
\begin{gather}
 \left.\Lax_\pm(z)\right|_{E=0}
 \doteq\frac{L_\pm}{1\mp z}\,,
 \label{eq:af-lambda-lax-e0}
\end{gather}
which is the standard lambda model Lax connection. In $(\tau,\sigma)$ coordinates, the spatial component needed below is
\begin{gather}
 \Lax_\sigma(\sigma,z)
 =\frac{\mathcal A_\sigma(\sigma)+z\mathcal B_\tau(\sigma)}{1-z^2}\,.
 \label{eq:af-lambda-spatial-lax}
\end{gather}
This expression will be the starting point for the Maillet calculation in Subsection~\ref{subsec:maillet}.

The same four current identities also generate infinitely many conserved currents. To see this, we use the generalized BIZZ construction of \cite{Bielli:2026ggs}, which begins with a nonzero constant $\beta_{\mathrm{BIZZ}}$ and potentials $\chi^{(\ell)}$ defined locally by
\begin{gather}
 \star J^{(\ell)}=\dd\chi^{(\ell)}\,.
 \label{eq:gbizz-potentials}
\end{gather}
For the currents $\mathcal A$ and $\mathcal B$, the first three members of the tower are
\begin{align}
 J^{(0)}&=\beta_{\mathrm{BIZZ}}\mathcal B\,,
 \notag\\
 J^{(1)}&=\star\mathcal A
 +\frac{1}{2\beta_{\mathrm{BIZZ}}}
 [\mathcal B,\chi^{(0)}]\,,
 \notag\\
 J^{(2)}&=\mathcal B
 +\frac{1}{2\beta_{\mathrm{BIZZ}}}
 [\star\mathcal A,\chi^{(0)}]
 +[\mathcal B,\chi^{(1)}]\,.
 \label{eq:gbizz-first-currents}
\end{align}
The remaining currents are obtained recursively from
\begin{gather}
 J^{(\ell)}=J^{(\ell-2)}
 +[\star\mathcal A,\chi^{(\ell-2)}]
 +[\mathcal B,\chi^{(\ell-1)}]\,,
 \qquad
 \dd(\star J^{(\ell)})\approx0\,,
 \qquad \ell\geq3\,.
 \label{eq:gbizz-recursion}
\end{gather}
It is instructive to check the first nontrivial current.  Using $\star J^{(0)}=\dd\chi^{(0)}$, conservation of $\mathcal B$, and the form identity $[\star\mathcal B,\star\mathcal B]=-[\mathcal B,\mathcal B]$, one obtains
\begin{align}
 \dd(\star J^{(1)})
 &=\dd\mathcal A
 +\frac{1}{2\beta_{\mathrm{BIZZ}}}
 \dd[\star\mathcal B,\chi^{(0)}]
 \notag\\
 &\approx-\frac12[\mathcal A,\mathcal A]
 +\frac12[\mathcal B,\mathcal B]
 \doteq0\,.
 \label{eq:gbizz-first-check}
\end{align}
At the next level, the mixed commutator identities remove the terms involving one copy of each current.  The same cancellation then repeats by the Jacobi identity, giving the conserved tower in \eqref{eq:gbizz-recursion}.  This is precisely the mechanism established in the general theorem of \cite{Bielli:2026ggs}; here its hypotheses are the four identities already derived from the AF-$\lambda$ equations.

The normalization $\beta_{\mathrm{BIZZ}}$ is arbitrary at the level of local conservation.  Its role can be seen from $\dd\chi^{(0)}=\star J^{(0)}=\beta_{\mathrm{BIZZ}}\star\mathcal B$. For a fixed choice of integration constants, write
\begin{gather}
 \chi^{(0)}=\beta_{\mathrm{BIZZ}}\widehat\chi^{(0)}\,,
 \qquad
 \dd\widehat\chi^{(0)}=\star\mathcal B\,.
\end{gather}
It follows that the explicit inverse power of $\beta_{\mathrm{BIZZ}}$ in the bilocal term cancels:
\begin{gather}
 \frac{1}{2\beta_{\mathrm{BIZZ}}}
 [\mathcal B,\chi^{(0)}]
 =\frac12[\mathcal B,\widehat\chi^{(0)}]\,.
\end{gather}
Thus rescaling $\beta_{\mathrm{BIZZ}}$ rescales $J^{(0)}$ and $\chi^{(0)}$, but does not change the relative normalization of the local and bilocal pieces of $J^{(1)}$.  Once Poisson brackets are introduced, however, the level-zero charge must generate the Lie algebra with its conventional normalization.  This fixes $\beta_{\mathrm{BIZZ}}$ uniquely in Subsection~\ref{subsec:yangian}.

When $E=0$, the two inputs $\mathcal{A}_\pm$, $\mathcal{B}_\pm$ coincide and \eqref{eq:gbizz-recursion} reduces to the original BIZZ tower of the lambda model. For nonzero $E$, the recursion involves both the flat current $\mathcal A$ and the conserved current $\mathcal B$. Locally, the conservation equations guarantee the existence of each potential $\chi^{(\ell)}$, by the Poincar\'e lemma.  The corresponding nonlocal charges additionally require a spatial base point and boundary conditions; we will specify these choices when we apply the Yangian theorem.

The spatial Lax matrix and the level-zero and level-one BIZZ charges depend on the same components, $\mathcal A_\sigma$ and $\mathcal B_\tau$. We now determine their equal-time algebra.

\subsection{Hamiltonian current algebra}
\label{subsec:current-algebra}

For the Hamiltonian analysis, we assume that $E$ is twice continuously differentiable.

Although the auxiliary fields $v_\pm$ are non-dynamical, in the Hamiltonian formulation they yield a constrained system, and therefore one should take these constraints into account and use the appropriate Dirac brackets when studying the Hamiltonian structure of the AF-$\lambda$ theory. However, as in other examples of auxiliary field sigma models \cite{Bielli:2024ach,Cesaro:2024ipq,Bielli:2026ggs}, it turns out that these additional constraints do not materially affect the canonical structure of the theory. We have collected the details of this argument in Appendix \ref{app:hamiltonian-constraints}; in short, in the gauge-fixed formulation used here, we need only assume that the remaining algebraic constraints form a second-class system and that the matrix $M$ displayed in equation (\ref{app:eq-regularity-block}) is invertible. Under this assumption, the inverse of the full constraint matrix has a zero block that makes the Dirac correction vanish for the canonical WZW currents; they therefore retain their Kac-Moody brackets (see \eqref{app:eq-constraint-block-inverse}). In the undeformed limit $E=0$, $\det M=\kappa^2(\lambda^{-2}-1)$ for each adjoint component, which is nonzero for $0<\lambda<1$ and $\kappa\neq0$. We therefore evaluate the canonical WZW brackets before imposing the algebraic constraints -- the latter then express the WZW currents in terms of $\mathcal A$ and $\mathcal B$.

Here and below, we call
\begin{align}\label{canonical_wzw_currents}
\mathscr J_+
=\kappa\bigl(-\partial_+g g^{-1} + ga_+g^{-1}-a_-\bigr) \, , \qquad 
\mathscr J_-
=\kappa\bigl(g^{-1}\partial_-g+g^{-1}a_-g-a_+\bigr)
\end{align}
the ``canonical WZW currents'' because they arise from the gauged-WZW part of the full lambda model at finite $\lambda$. More precisely, (\ref{canonical_wzw_currents}) are the expressions for these currents in the Lagrangian formulation. In the Hamiltonian analysis below, we use the same currents as functions of $g$ and its conjugate momentum $\Pi_g$, as explained in Appendix \ref{app:hamiltonian-constraints}. Although we use this convenient choice of variables, we stress that we work with the full auxiliary field lambda model and are not taking a WZW limit. The Kac-Moody algebra of $\mathscr{J}_\pm$ only provides the starting point for deriving the brackets of $\mathcal A_\sigma$ and $\mathcal B_\tau$.

Let $T_A$ be a basis of $\mathfrak g$, with $[T_A,T_B]=f_{AB}{}^C T_C$, and let $\gamma_{AB}=\tr(T_AT_B)$ be the invariant form.  Writing $\delta' ( \sigma - \sigma' ) =\partial_\sigma\delta(\sigma-\sigma')$, the canonical WZW currents $\mathscr J_\pm$ obey \cite{Papadopoulos:1992rs,Schmidtt:2017bqy}
\begin{align}
 \{\mathscr J_\pm^A(\sigma),\mathscr J_\pm^B(\sigma')\}_D
 &={}
 f^{AB}{}_C\mathscr J_\pm^C(\sigma)\delta(\sigma-\sigma')
 \mp2\kappa\gamma^{AB}\delta' ( \sigma - \sigma' ) \,,
 \notag\\
 \{\mathscr J_+^A(\sigma),\mathscr J_-^B(\sigma')\}_D&=0\,.
 \label{eq:wzw-km-algebra}
\end{align}
Here the WZW level is $k_{\mathrm{WZW}}=4\pi\kappa$. Only after evaluating the canonical brackets involving $\mathscr{J}_\pm$ do we impose the algebraic constraints; on the constraint surface they give
\begin{gather}
 \mathscr J_\pm
 \simeq
 \kappa\bigl(\lambda a_\pm-q_\lambda v_\pm-a_\mp\bigr)\,,
 \qquad
 q_\lambda=\lambda-\lambda^{-1}\,.
 \label{eq:wzw-reduced-constraints}
\end{gather}
We use $\simeq$ for equality on the Hamiltonian constraint surface, distinguishing it from $\doteq$ and $\approx$.

Substituting the inverse relations in \eqref{eq:split-current-inverse} into \eqref{eq:wzw-reduced-constraints} now gives
\begin{gather}
 {
 \mathscr J_\pm
 \simeq
 c_\lambda\mathcal B_\tau
 \pm d_\lambda\mathcal A_\sigma\,,}
 \qquad
 c_\lambda=\frac{\kappa(1-\lambda^2)}{2\lambda}\,,
 \qquad
 d_\lambda=\frac{\kappa(1+\lambda)^2}{2\lambda}\,.
 \label{eq:wzw-split-current-relation}
\end{gather}
The coefficients follow by collecting the sum $\mathcal B_++\mathcal B_-=2\mathcal B_\tau$ and the difference $\mathcal A_+-\mathcal A_-=2\mathcal A_\sigma$.  In particular, every derivative of the interaction function that appeared in the auxiliary equations has dropped out.  The inverse relation is
\begin{gather}
 \mathcal B_\tau
 \simeq\frac{\mathscr J_++\mathscr J_-}{2c_\lambda}\,,
 \qquad
 \mathcal A_\sigma
 \simeq\frac{\mathscr J_+-\mathscr J_-}{2d_\lambda}\,.
 \label{eq:wzw-split-current-inverse}
\end{gather}
This is the main structural simplification in the Hamiltonian analysis.  The chosen interaction function $E$ affects the solution for $a_\pm$ and $v_\pm$ and hence the Hamiltonian, but the two combinations entering the spatial Lax matrix are fixed linear combinations of $\mathscr J_+$ and $\mathscr J_-$. It also explains why the two opposite affine levels survive the deformation: the auxiliary fields alter the realization of the currents, not the WZW symplectic form from which their central terms originate.

To derive the resulting current algebra, we first define the sum $\mathscr S=\mathscr J_++\mathscr J_-$ and difference $\mathscr D=\mathscr J_+-\mathscr J_-$.  Equation \eqref{eq:wzw-km-algebra} then yields
\begin{align}
 \{\mathscr S^A ( \sigma ) ,\mathscr S^B ( \sigma' ) \}_D
 &=f^{AB}{}_C\mathscr S^C ( \sigma ) \delta ( \sigma - \sigma' ) \,,
 \notag\\
 \{\mathscr S^A ( \sigma ) ,\mathscr D^B ( \sigma' ) \}_D
 &=f^{AB}{}_C\mathscr D^C ( \sigma ) \delta ( \sigma - \sigma' ) 
 -4\kappa\gamma^{AB}\delta' ( \sigma - \sigma' ) \,,
 \label{eq:wzw-sum-difference-algebra}\\
 \{\mathscr D^A ( \sigma ) ,\mathscr D^B ( \sigma' ) \}_D
 &=f^{AB}{}_C\mathscr S^C ( \sigma ) \delta ( \sigma - \sigma' ) \,.
 \notag
\end{align}
The affine terms cancel in the first and third lines because the two Kac-Moody levels have opposite signs.  In the mixed bracket, the minus sign in $\mathscr D$ reverses the contribution from $\mathscr J_-$, so the two derivative terms add to $-4\kappa\gamma^{AB}\delta' ( \sigma - \sigma' )$.  Dividing $\mathscr S$ and $\mathscr D$ by the coefficients in \eqref{eq:wzw-split-current-inverse}, we obtain
\begin{align}
 \{\mathcal B_\tau^A(\sigma),\mathcal B_\tau^B(\sigma')\}_D
 &={}
 \frac{1}{2c_\lambda}f^{AB}{}_C
 \mathcal B_\tau^C(\sigma)\delta ( \sigma - \sigma' ) \,,
 \notag\\
 \{\mathcal B_\tau^A(\sigma),\mathcal A_\sigma^B(\sigma')\}_D
 &={}
 \frac{1}{2c_\lambda}f^{AB}{}_C
 \mathcal A_\sigma^C(\sigma)\delta ( \sigma - \sigma' ) 
 -\frac{\kappa}{c_\lambda d_\lambda}\gamma^{AB}\delta' ( \sigma - \sigma' ) \,,
 \label{eq:split-current-algebra}\\
 \{\mathcal A_\sigma^A(\sigma),\mathcal A_\sigma^B(\sigma')\}_D
 &={}
 \frac{c_\lambda}{2d_\lambda^2}f^{AB}{}_C
 \mathcal B_\tau^C(\sigma)\delta ( \sigma - \sigma' ) \,.
 \notag
\end{align}
No field equation beyond the constraint relation \eqref{eq:wzw-split-current-relation} enters this bracket calculation.  The result is an invertible linear image of two commuting Kac-Moody algebras and therefore satisfies the Jacobi identities.  It also shows directly that the only non-ultralocal term occurs in the mixed bracket.

It is useful at this point to summarize the two results of \cite{Bielli:2026ggs} that will be used below.  The first is the generalized BIZZ theorem already applied in Subsection~\ref{subsec:lax-integrability}: a flat current $\mathcal A$ and a conserved current $\mathcal B$ satisfying the two commutator identities in \eqref{eq:split-commutator-identities} generate the conserved tower beginning with \eqref{eq:gbizz-first-currents}.  This theorem guarantees conservation of the nonlocal currents, but does not by itself determine the Poisson algebra of their charges.

Theorem 3.1 of \cite{Bielli:2026ggs} supplies the additional Hamiltonian input needed for a Yangian.  Since the Dirac bracket is the Poisson bracket on the reduced phase space, we may apply this result with its Poisson brackets understood here as the reduced brackets $\{\ ,\ \}_D$.  In notation adapted to the present currents, the theorem considers the equal-time algebra
\begin{align}
 \{\mathcal B_\tau^A(\sigma),\mathcal B_\tau^B(\sigma')\}_D
 &={}
 m_1 f^{AB}{}_C\mathcal B_\tau^C(\sigma)\delta ( \sigma - \sigma' )
 +m_2\gamma^{AB}\delta' ( \sigma - \sigma' ) \, ,
 \notag\\
 \{\mathcal B_\tau^A(\sigma),\mathcal A_\sigma^B(\sigma')\}_D
 &={}
 m_3 f^{AB}{}_C\mathcal A_\sigma^C(\sigma)\delta ( \sigma - \sigma' )
 +m_4 C^{AB}(\sigma')\delta' ( \sigma - \sigma' ) \, ,
 \notag\\
 \{\mathcal A_\sigma^A(\sigma),\mathcal A_\sigma^B(\sigma')\}_D
 &={}
 f^{AB}{}_C
 \bigl(
 m_5\mathcal A_\sigma^C(\sigma)
 +m_6\mathcal B_\tau^C(\sigma)
 \bigr)\delta ( \sigma - \sigma' )
 +m_7\gamma^{AB}\delta' ( \sigma - \sigma' ) \, .
 \label{eq:generalized-yangian-current-algebra}
\end{align}
Here $m_1,\ldots,m_7$ are constant coefficients labeling the structures retained in this general current algebra, while $C^{AB}$ is a symmetric tensor which may in general depend on the fields.  Thus the $m_i$ are not additional couplings of the AF-$\lambda$ model.

For a finite-dimensional simple Lie algebra,\footnote{Throughout this paper, as in \cite{Bielli:2026ggs}, we exclude the special $\mathfrak{sl}(2)$ case, including the real algebras $\mathfrak{su}(2)$ and $\mathfrak{sl}(2,\mathbb R)$. More precisely, the complexified algebra must have no $\mathfrak{sl}(2,\mathbb C)$ factor (type $A_1$). For $\mathfrak{sl}(2)$, the first Serre relation is identically satisfied, so the second must be checked separately. We leave this to future work.} the first two generalized BIZZ charges obey the classical Yangian relations when the previously arbitrary normalization is fixed according to $\beta_{\mathrm Y}:=\beta_{\mathrm{BIZZ}}=m_1^{-1}$, when $m_3=m_1\neq0$, and when the $C^{AB}$-dependent contribution vanishes after the antisymmetrization in the Serre relation, as specified in \eqref{eq:yangian-integrated-obstruction}. The last condition is automatic when $C^{AB}=\gamma^{AB}$.  The same current-algebra parametrization also controls the Maillet calculation: in the case $C^{AB}=\gamma^{AB}$, the four relations collected in \eqref{eq:maillet-conditions} ensure that the spatial Lax matrix has a Maillet bracket with field-independent $r/s$ kernels proportional to the quadratic Casimir.

Matching \eqref{eq:split-current-algebra} with \eqref{eq:generalized-yangian-current-algebra}, we can write the nonzero coefficients as
\begin{gather}
 m_1=m_3=\frac{1}{2c_\lambda}\,,
 \qquad
 m_4=-\frac{\kappa}{c_\lambda d_\lambda}\,,
 \qquad
 m_6=\frac{c_\lambda}{2d_\lambda^2}\,.
 \label{eq:mi-current-map-form}
\end{gather}
The remaining coefficients multiply structures absent from \eqref{eq:split-current-algebra}.  The complete set and the symmetric tensor are therefore
\begin{gather}
 {
 \begin{gathered}
 m_1=m_3=\frac{\lambda}{\kappa(1-\lambda^2)}\,,
 \qquad
 m_2=m_5=m_7=0\,,\\[2mm]
 m_4=-\frac{4\lambda^2}
 {\kappa(1-\lambda)(1+\lambda)^3}\,,
 \qquad
 m_6=\frac{\lambda(1-\lambda)}
 {\kappa(1+\lambda)^3}\,,
 \qquad
 C^{AB}=\gamma^{AB}\,.
 \end{gathered}}
 \label{eq:af-lambda-mi}
\end{gather}
A common denominator also gives the useful identity
\begin{gather}
 m_4=m_6-m_1\,.
 \label{eq:mi-identity}
\end{gather}
The auxiliary interaction changes the dynamics but leaves the equal-time algebra \eqref{eq:split-current-algebra} unchanged. As a check, setting $E=0$ gives $\mathcal A_\pm\doteq\mathcal B_\pm\doteq L_\pm$ but leaves every coefficient in \eqref{eq:af-lambda-mi} untouched.  We have thus recovered the ordinary lambda model current algebra in precisely the normalization used for the Lax connection.

We will now use different parts of these coefficients for two purposes.  The Yangian theorem depends on the normalization of the $\{\mathcal B_\tau,\mathcal B_\tau\}_D$ bracket, the adjoint action in the mixed bracket, and the invariant tensor multiplying $\delta' ( \sigma - \sigma' )$.  The Maillet calculation uses all three brackets because the spatial Lax matrix is a spectral-parameter-dependent combination of $\mathcal A_\sigma$ and $\mathcal B_\tau$.  No further Hamiltonian reduction is needed in either calculation.

\subsection{Classical Yangian symmetry}
\label{subsec:yangian}

We first apply this current algebra to the nonlocal charges. The classical Yangian symmetry of the ordinary lambda model was derived directly from its nonlocal charges in \cite{Itsios:2014lca}. In the first realization, a Yangian is generated by a level-zero copy of $\mathfrak g$ and a level-one generator transforming in its adjoint representation, subject to the Serre relations \cite{Drinfeld:1985rx,MacKay:1992he}.  The theorem of \cite{Bielli:2026ggs} obtains these generators from the first two generalized BIZZ currents, provided that their equal-time algebra has the appropriate normalization and tensor structure. In our conventions the relevant charge densities are
\begin{gather}
 J_\tau^{(0)}=\beta_{\mathrm Y}\mathcal B_\tau\,,
 \qquad
 J_\tau^{(1)}=\mathcal A_\sigma
 +\frac{1}{2\beta_{\mathrm Y}}
 [\mathcal B_\tau,\chi^{(0)}]\,,
 \qquad
 \mathcal B_\tau=\frac{1}{\beta_{\mathrm Y}}
 \partial_\sigma\chi^{(0)}\,.
 \label{eq:yangian-current-densities}
\end{gather}
The first term in $J_\tau^{(1)}$ is local, while the potential $\chi^{(0)}$ makes the second term bilocal after solving its defining equation.  We choose its base point at the left end of the line, so that $\chi^{(0)}(-\infty)=0$.  Formally, after the ordered cutoff has been removed,
\begin{gather}
 \chi^{(0)}(\sigma)
 =\beta_{\mathrm Y}\int_{-\infty}^{\sigma}d\sigma'\,
 \mathcal B_\tau(\sigma')\,.
 \label{eq:yangian-potential}
\end{gather}
The corresponding charges on the spatial line are
\begin{gather}
 Q^{(\ell)A}=\int_{-\infty}^{+\infty}
 d\sigma\,J_\tau^{(\ell)A}\,,
 \qquad \ell=0,1\,.
 \label{eq:yangian-charges}
\end{gather}
We assume that the currents fall off sufficiently rapidly for the charge integrals and bilocal products to converge and for the spatial flux to vanish. The derivative terms in \eqref{eq:split-current-algebra} are treated with the six-endpoint ordered prescription \eqref{eq:yangian-regulator-order}: the cutoffs at the positive end obey $\sigma_6<\sigma_4<\sigma_2$, while those at the negative end obey $\sigma_5<\sigma_3<\sigma_1$, before all six are sent to infinity.  This ordering places the endpoints of the local and bilocal parts of $Q^{(1)}$ inside those of $Q^{(0)}$. Integrating the differentiated delta function over the larger interval first then removes the resulting boundary distributions.  Its finite-cutoff form is given in Appendix~\ref{app:yangian-maillet}.

The normalization and first coefficient condition follow immediately from \eqref{eq:af-lambda-mi}:
\begin{gather}
 {
 \beta_{\mathrm Y}=m_1^{-1}
 =\frac{\kappa(1-\lambda^2)}{\lambda}
 =2c_\lambda\,,}
 \qquad
 m_3=m_1\neq0\,,
 \qquad
 \alpha_{\mathrm Y}=m_1^2
 =\frac{\lambda^2}{\kappa^2(1-\lambda^2)^2}\,.
 \label{eq:yangian-normalization}
\end{gather}
The first equality fixes the Lie bracket of the level-zero charge.  Indeed, \eqref{eq:wzw-split-current-inverse} gives $\beta_{\mathrm Y}\mathcal B_\tau=\mathscr J_++\mathscr J_-$, so $Q^{(0)}$ is the integral of the diagonal WZW current.  The equality $m_3=m_1$ then ensures that the local and bilocal pieces of $Q^{(1)}$ both transform in the adjoint representation with the same coefficient.  The quantity $\alpha_{\mathrm Y}$ is the Yangian deformation parameter in these conventions.

The normalization can also be checked directly.  Integrating the first bracket in \eqref{eq:split-current-algebra} gives
\begin{align}
 \{Q^{(0)A},Q^{(0)B}\}_D
 &=\beta_{\mathrm Y}^2m_1
 f^{AB}{}_C\int d\sigma\,
 \mathcal B_\tau^C(\sigma)
 \notag\\
 &=\beta_{\mathrm Y}m_1
 f^{AB}{}_CQ^{(0)C}\,.
 \label{eq:yangian-level-zero-check}
\end{align}
Thus $\beta_{\mathrm Y}=m_1^{-1}$ produces the standard Lie bracket.  In the mixed charge bracket, $m_3=m_1$ makes the local $\mathcal A_\sigma$ term transform with the same coefficient, while the Leibniz rule applies the adjoint action to both entries of the bilocal term. The ordered regulator removes the surface terms generated by $m_4\delta' ( \sigma - \sigma' )$. These observations establish the first two Yangian brackets. It remains to check the tensor condition entering the Serre relation.

The remaining Poisson bracket, $\{Q^{(1)},Q^{(1)}\}_D$, contains both local and trilocal terms.  After contraction with $f_D{}^{[AB}$ and antisymmetrization in $A,B,C$, the trilocal contribution reorganizes into the cubic product of level-zero charges in \eqref{eq:yangian-serre}.  Terms built from the derivative part of the mixed current bracket either vanish under the ordered endpoint prescription or assemble into the tensor \eqref{eq:yangian-integrated-obstruction}.

Since $C^{AB}=\gamma^{AB}$, invariance of the quadratic Casimir gives
\begin{gather}
 \mathcal B_\tau^D
 \bigl(f_D{}^{AC}\delta_C{}^B-f_D{}^{BC}\delta_C{}^A\bigr)
 =2\mathcal B_\tau^D f_D{}^{AB}\,.
 \label{eq:yangian-centrality}
\end{gather}
The tensor appearing in the theorem is
\begin{gather}
 \mathcal V_1^{CD}
 =\int_{-\infty}^{+\infty}d\sigma\,
 \mathcal B_\tau^E
 \bigl(f_E{}^{CF}C_F{}^D-f_E{}^{DF}C_F{}^C\bigr)\,,
 \qquad
 f_D{}^{[AB}\mathcal V_1^{C]D}=0\,.
 \label{eq:yangian-integrated-obstruction}
\end{gather}
Equation \eqref{eq:yangian-centrality} puts the integrand in the form $2\mathcal B_\tau^E f_E{}^{CD}$ pointwise. Its antisymmetrized contraction with a second structure constant therefore vanishes by the Lie algebra Jacobi identity.  This completes the model-dependent part of the theorem's hypotheses.

The theorem now gives the level-zero and level-one brackets
\begin{gather}
 \{Q^{(0)A},Q^{(0)B}\}_D
 =f^{AB}{}_C Q^{(0)C}\,,
 \qquad
 \{Q^{(0)A},Q^{(1)B}\}_D
 =f^{AB}{}_C Q^{(1)C}\,,
 \label{eq:yangian-level-zero-one}
\end{gather}
together with the first Serre relation
\begin{align}
 f_D{}^{[AB}\{Q^{(1)C]},Q^{(1)D}\}_D
 &={}
 \frac{\alpha_{\mathrm Y}^2}{12}
 f^A{}_{PI}f^B{}_{QJ}f^C{}_{RK}f^{IJK}
 Q^{(0)P}Q^{(0)Q}Q^{(0)R}\,,
 \notag\\[-1mm]
 \frac{\alpha_{\mathrm Y}^2}{12}
 &=\frac{m_1^4}{12}
 =\frac{\lambda^4}
 {12\kappa^4(1-\lambda^2)^4}\,.
 \label{eq:yangian-serre}
\end{align}
The factor in the last line is worth noting: the Serre coefficient is $\alpha_{\mathrm Y}^2/12=m_1^4/12$, rather than $m_1^2/12$.  This normalization follows from choosing $\beta_{\mathrm Y}=m_1^{-1}$ in the level-zero current and agrees with the first-realization convention of \cite{Bielli:2026ggs}.

The generalized BIZZ equations also establish conservation.  At either of the first two levels,
\begin{gather}
 \partial_\tau Q^{(\ell)}
 =J_\sigma^{(\ell)}(+\infty)
  -J_\sigma^{(\ell)}(-\infty)=0\,,
 \qquad \ell=0,1\,,
 \label{eq:yangian-charge-conservation}
\end{gather}
where the last equality follows from the stated falloff. The first Serre relation then implies the second \cite{Drinfeld:1985rx}.

We conclude that the AF-$\lambda$ charges therefore generate the classical Yangian $Y_{\mathrm C}(\mathfrak g)$ by Theorem 3.1 of \cite{Bielli:2026ggs}. The result combines two pieces of information: the Lagrangian equations give the conserved generalized BIZZ currents, and the unchanged WZW algebra fixes their equal-time brackets.  The auxiliary interaction is present in the time evolution that establishes conservation, but not in the Yangian structure constants or in the normalization \eqref{eq:yangian-normalization}.

\subsection{Maillet algebra and twist function}
\label{subsec:maillet}

We now apply the same current algebra to the spatial Lax matrix.  Because the mixed bracket in \eqref{eq:split-current-algebra} contains a derivative of a delta function, the model is non-ultralocal.  The appropriate framework is the $r/s$ algebra introduced by Maillet \cite{MAILLET198654,Maillet:1985ec}. Section 4.4 of \cite{Bielli:2026ggs} gives conditions for the current algebra \eqref{eq:generalized-yangian-current-algebra} to induce a Maillet bracket for the Lax matrix. For $C^{AB}=\gamma^{AB}$, as in the present model, these conditions are
\begin{gather}
 m_3=m_1\,,
 \qquad
 m_5+m_7=0\,,
 \qquad
 m_2+m_5=0\,,
 \qquad
 m_3+m_4-m_6=0\,.
 \label{eq:maillet-conditions}
\end{gather}
All four are satisfied by \eqref{eq:af-lambda-mi}; the last reduces to $m_4=m_6-m_1$.  The first follows because the Lie terms in the first and mixed brackets have the same coefficient, and the next two follow from the absent structures in \eqref{eq:split-current-algebra}.  The fourth matches the ultralocal and derivative terms using the same pair of scalar kernels.

To see why these conditions arise, use the three current brackets to compute the bracket of $\Lax_\sigma=(\mathcal A_\sigma+z\mathcal B_\tau)/(1-z^2)$.  The coefficient of $f^{AB}{}_C\mathcal A_\sigma^C\delta ( \sigma - \sigma' )$ is proportional to $m_3(z+z')+m_5$, the coefficient of $f^{AB}{}_C\mathcal B_\tau^C\delta ( \sigma - \sigma' )$ to $m_1zz'+m_6$, and the derivative term to $m_4(z+z')+m_2zz'+m_7$.  A single Casimir-valued kernel must reproduce all three polynomials.  Equation \eqref{eq:maillet-conditions} is the compatibility condition for doing so at generic $z$ and $z'$.

Let $C_{12}=\gamma^{AB}T_A\otimes T_B$, and write $X_1=X\otimes\id$ and $X_2=\id\otimes X$.  We use the convention
\begin{align}
 \{\Lax_{\sigma,1}(\sigma,z),
    \Lax_{\sigma,2}(\sigma',z')\}_D
 &= [r_{12}(z,z'),\Lax_{\sigma,1}(\sigma,z)]\delta ( \sigma - \sigma' )
 -[r_{21}(z',z),\Lax_{\sigma,2}(\sigma,z')]\delta ( \sigma - \sigma' )
 \nonumber \\
 &\qquad -s_{12}(z,z')\delta' ( \sigma - \sigma' )\,,
 \label{eq:af-lambda-maillet-bracket}\\
 s_{12}(z,z')&=r_{12}(z,z')+r_{21}(z',z)\,.
 \notag
\end{align}
This fixes both the sign of the $\delta' ( \sigma - \sigma' )$ term and the normalization of $s_{12}$.  Invariance of the quadratic Casimir converts the commutators in \eqref{eq:af-lambda-maillet-bracket} into the component structures appearing in \eqref{eq:split-current-algebra}.  A field-independent ansatz $r_{12}(z,z')=f(z,z')C_{12}$ then reduces the calculation to two scalar matching equations.  Solving them and checking the symmetric part of $f$ gives
\begin{align}
 r_{12}(z,z')
 &=\frac{\lambda}{\kappa(1-\lambda^2)}
 \frac{z_\lambda^2-z'^2}{(z-z')(1-z'^2)}C_{12}\,, \nonumber \\
 s_{12}(z,z')
 &=\frac{4\lambda^2}
 {\kappa(1-\lambda)(1+\lambda)^3}
 \frac{z+z'}{(1-z^2)(1-z'^2)}C_{12}\,, \nonumber \\
 z_\lambda &= \frac{1-\lambda}{1+\lambda}\,.
 \label{eq:af-lambda-rs}
\end{align}
Appendix~\ref{app:yangian-maillet} gives the component bracket and the scalar matching, including the sign of the non-ultralocal term in \eqref{eq:af-lambda-maillet-bracket}.

Several features of \eqref{eq:af-lambda-rs} are already visible from the current algebra.  The pole at $z=z'$ is the usual rational Casimir pole.  The nonzero $s$-kernel is determined by $m_4$, and therefore by the difference of the two Kac-Moody central terms in \eqref{eq:wzw-km-algebra}.  Both kernels are independent of the worldsheet fields, so the standard symmetrized transport-matrix construction applies.  Finally, all dependence on the auxiliary interaction has disappeared because the calculation uses only the linear WZW relation \eqref{eq:wzw-split-current-relation}.

The Jacobi identity follows from the Dirac bracket \eqref{eq:split-current-algebra}. The nonzero symmetric kernel records the non-ultralocal term in the mixed current bracket.

The first kernel in \eqref{eq:af-lambda-rs} has the twist-function form
\begin{gather}
 r_{12}(z,z')
 =\frac{C_{12}}{z-z'}\,\varphi_\lambda^{-1}(z')\,,
 \qquad
 {
 \varphi_\lambda(z)
 =\frac{\kappa(1-\lambda^2)}{\lambda}
 \frac{1-z^2}{z_\lambda^2-z^2}\,.}
 \label{eq:af-lambda-twist}
\end{gather}
This is the standard lambda-model twist function in our spectral and Casimir conventions \cite{Georgiou:2019plp}. To see how its parameters arise, note that
\begin{gather}
 m_6=m_1z_\lambda^2\,,
 \qquad
 \varphi_\lambda(z)=\frac{1-z^2}{m_6-m_1z^2}\,.
 \label{eq:twist-site}
\end{gather}
The prefactor in \eqref{eq:af-lambda-twist} is therefore $m_1^{-1}=\beta_{\mathrm Y}$.  It is fixed by the affine levels rather than by a separate rescaling of the spectral parameter.

The poles of $\varphi_\lambda$ also identify the two WZW currents inside the Lax matrix.  Since $c_\lambda/d_\lambda=z_\lambda$ and $d_\lambda(1-z_\lambda^2)=2\kappa$, evaluation of \eqref{eq:af-lambda-spatial-lax} gives
\begin{gather}
 {
 \Lax_\sigma(z_\lambda)=\frac{\mathscr J_+}{2\kappa}\,,
 \qquad
 \Lax_\sigma(-z_\lambda)=-\frac{\mathscr J_-}{2\kappa}\,,}
 \label{eq:lax-at-twist-poles}
\end{gather}
while the residues of $\varphi_\lambda(z)\,d z$ are
\begin{gather}
 {
 \operatorname*{res}_{z=z_\lambda}
 \varphi_\lambda(z)\,d z=-2\kappa\,,
 \qquad
 \operatorname*{res}_{z=-z_\lambda}
 \varphi_\lambda(z)\,d z=+2\kappa\,.}
 \label{eq:twist-residues}
\end{gather}
The two pole values and opposite residues reproduce the WZW currents and their Kac-Moody levels; Appendix~\ref{app:yangian-maillet} gives the short calculation. Since $0<z_\lambda<1$ in the parameter range used here, the Lax matrix is regular at the poles $z=\pm z_\lambda$ of the twist function; its own poles remain at $z=\pm1$.

Let us finish by extracting the conserved commuting quantities.  For two spatial points $\sigma_1$ and $\sigma_2$, define the transport matrix
\begin{gather}
 \mathcal T(\sigma_2,\sigma_1;z)
 =\operatorname{Pexp}\!\left(
 -\int_{\sigma_1}^{\sigma_2}d\sigma\,
 \Lax_\sigma(\tau,\sigma;z)\right)\,.
 \label{eq:transport}
\end{gather}
The zero-curvature equation \eqref{eq:af-lambda-lax-curvature} implies
\begin{align}
 \partial_\tau\mathcal T(\sigma_2,\sigma_1;z)
 =-\Lax_\tau(\sigma_2,z)
   \mathcal T(\sigma_2,\sigma_1;z)
 +\mathcal T(\sigma_2,\sigma_1;z)
  \Lax_\tau(\sigma_1,z)\,.
 \label{eq:transport-time-evolution}
\end{align}
On a circle, periodicity makes the time evolution of the monodromy $\mathcal M(z)$ a conjugation.  Explicitly, for a circle of circumference $R$ and base point $\sigma_0$ we take
\begin{gather}
 \mathcal M(z)=
 \mathcal T(\sigma_0+R,\sigma_0;z)\,,
 \qquad
 \partial_\tau\mathcal M(z)
 =[\mathcal M(z),\Lax_\tau(\sigma_0,z)]\,.
 \label{eq:monodromy-evolution}
\end{gather}
On the line, one instead uses $\mathcal M(z)=\mathcal T(+\infty,-\infty;z)$ and imposes endpoint behavior which removes the two terms in \eqref{eq:transport-time-evolution}.  In either case cyclicity of the trace shows that $\tr\mathcal M(z)^p$ is conserved for every positive integer $p$.

Conservation and involution use different ingredients.  The preceding transport equation follows from zero curvature, whereas mutual involution follows from the equal-time Maillet algebra.  Using the symmetric point-splitting prescription for coincident transport endpoints \cite{MAILLET198654,Maillet:1985ec}, one obtains
\begin{gather}
 {
 \left\{
 \tr\mathcal M(z)^p,
 \tr\mathcal M(z')^r
 \right\}_M=0\,,
 \qquad p,r\in\mathbb N\,.}
 \label{eq:monodromy-involution}
\end{gather}
Here $\{\ ,\ \}_M$ denotes the symmetrized coincident-endpoint bracket.  The result applies on the circle or the line whenever the corresponding monodromy and boundary conditions are well defined.

The current algebra, Maillet kernels, and twist function are therefore the same for every interaction function $E$ for which the matrix in Appendix~\ref{app:hamiltonian-constraints} is invertible. When $E=0$, this statement reduces to the familiar Hamiltonian structure of the lambda model.  For general $E$, the solutions of the algebraic field equations change, but $\mathcal A_\sigma$ and $\mathcal B_\tau$ remain the same linear combinations of the canonical WZW currents. Consequently, the Lax matrix still obeys \eqref{eq:lax-at-twist-poles}, and the twist function is unchanged.  This is why the auxiliary interaction can change the equations of motion while preserving both the Yangian algebra and the commuting monodromy invariants.
\section{Conclusions}
\label{sec:conclusion}

In this work, we have constructed an auxiliary field deformation of the lambda model.  The deformation acts on the gauge-fixed principal chiral model sector and is parametrized by an arbitrary differentiable interaction $E(\nu_2,\ldots,\nu_N)$.  When $E=0$, the auxiliary equations set $v_\pm=a_\pm$ and the standard lambda action is recovered exactly. For nonzero $E$, the equations of motion imply that $\mathcal A$ is flat and $\mathcal B$ is conserved. The auxiliary field equations also imply the two algebraic identities needed by the generalized BIZZ construction.  These currents give a Lax connection and an infinite tower of conserved nonlocal currents.  The shifted formulation also gives a simple perturbative interpretation: an interaction $E=\varepsilon\nu_n$ produces a bilinear of spin-$n$ currents of the ordinary lambda model to first order in $\varepsilon$. Eliminating the gauge fields gives the sigma model action, whose non-Abelian T-dual limit is the higher-spin auxiliary field deformation of the T-dual PCM. The current variables remain finite in this limit and the Lax connection reduces to that of the T-dual theory, so the auxiliary field construction commutes with the lambda-to-NATD limit. At $E=0$, the two currents coincide with the usual lambda-model current and the Lax connection reduces to its standard rational form. The Lax and generalized BIZZ constructions use the commutator identities implied by conjugation invariance of $E$, without requiring its detailed form.

We have also determined the equal-time algebra underlying these conserved quantities.  The main assumption in this calculation is that, in the gauge-fixed formulation used here, the algebraic constraint matrix for $a_\pm$ and $v_\pm$ is invertible.  The canonical WZW currents then retain their two commuting Kac-Moody algebras, and an interaction-independent linear relation transfers these brackets to $\mathcal B_\tau$ and $\mathcal A_\sigma$.  The resulting coefficients satisfy the hypotheses of the Yangian theorem, so the level-zero and level-one charges generate the classical Yangian on the spatial line.  The same current algebra places the spatial Lax matrix in Maillet $r/s$ form.  Its twist function and the two Kac-Moody levels are precisely those of the lambda model; with the usual symmetric prescription, the corresponding monodromy invariants are in involution.  Thus the auxiliary interaction changes the dynamics while leaving the affine current algebra, and consequently the classical Yangian and Maillet data, unchanged. This provides a direct example in which a broad family of auxiliary field interactions preserves the spectral and nonlocal symmetry structures of the original integrable theory.

There are several natural directions in which to extend these results.  The first concerns quantization.  One could begin with a point-split definition of the level-one charge and determine the local counterterms compatible with the global symmetry \cite{Luscher:1977uq}.  For polynomial choices of $E$, a one-loop calculation could test whether renormalization preserves the family of interactions $E(\nu_2,\ldots,\nu_N)$ and whether an anomaly obstructs conservation of the nonlocal charges at this order. This would be a first test of whether the classical Yangian survives quantization and could clarify how to organize the necessary counterterms.

A second question is how the generalized BIZZ charges are encoded in the monodromy matrix.  Expanding the monodromy near a convenient spectral point should allow one to compare its first coefficients with $Q^{(0)}$ and $Q^{(1)}$, including the bilocal kernel and its normalization.  It would also be useful to develop the analogous construction on the cylinder, where the base point and the periodic Green function must be treated explicitly. Such an analysis may clarify the relation between the line and circle descriptions.  It may also identify which expansion of the monodromy gives the most natural Drinfeld generators.

Finally, the persistence of the lambda-model twist function suggests a geometric explanation.  In an affine-Gaudin or four-dimensional Chern-Simons description \cite{Vicedo:2019dej,Delduc:2019whp}, one would like to identify a boundary or defect interaction that reproduces the algebraic $v_\pm$ equations while leaving the two affine sites and their levels fixed.  This perspective could explain the independence of the Maillet data from $E$ without a model-by-model calculation and may give a systematic route to auxiliary field deformations of anisotropic, coset, supercoset, and multi-site lambda models \cite{Sfetsos:2014lla,Hollowood:2014rla,Hollowood:2014qma,Georgiou:2017oly}.

\section*{Acknowledgments}

We are grateful to Daniele Bielli, Parita Shah, and Gabriele Tartaglino-Mazzucchelli for productive conversations related to this work, and especially to Jitendra Pal for collaboration in the initial stages of this project. We also acknowledge the language models ChatGPT-5.6 Sol and ChatGPT-6 Astra, which were used in this research. The authors take full responsibility for the accuracy of this work, having reviewed, verified, and approved all AI-generated content. C.\,F. is supported by the National Science Foundation under Cooperative Agreement PHY-2019786 (the NSF AI Institute for Artificial Intelligence and Fundamental Interactions).

\appendix
\section{Details of the equations of motion}
\label{app:eoms-lax}

In this appendix, we study the equations of motion for the auxiliary field deformed lambda model and derive the corresponding properties of the currents $\mathcal{A}$ and $\mathcal{B}$. We begin with the auxiliary field identity, then derive the group and gauge equations, and finally show how their sum and difference produce the flatness and conservation equations used in the main text.

Recall that the interaction function $E$ depends on the trace invariants $\nu_n$ defined in \eqref{eq:af-invariants}, and that the associated derivatives of $E$ are defined in \eqref{eq:delta-definition}. These definitions automatically imply certain convenient commutator identities. Indeed, an infinitesimal conjugation of $v_-$, with $v_+$ held fixed, gives
\begin{gather}
 0=\delta_XE
 =\tr\bigl(\Delta_+[X,v_-]\bigr)
 =\tr\bigl([v_-,\Delta_+]X\bigr)\,,
 \qquad X\in\mathfrak g\,.
 \label{app:eq-gradient-invariance-proof}
\end{gather}
Nondegeneracy of the invariant form implies $[\Delta_+,v_-]=0$; conjugating $v_+$ instead gives $[\Delta_-,v_+]=0$.  This proves the commutator identity \eqref{eq:delta-commutator} without using any differential equation of motion.

We next vary the action \eqref{eq:af-lambda-action}.  Its auxiliary field variation is
\begin{align}
 \delta_vS_{\mathrm{AF}\text{-}\lambda}
 =\kappa q_\lambda\int d^2x\,\tr\Bigl(&
 \delta v_-\bigl(a_+-v_+-\Delta_+\bigr) +\delta v_+\bigl(a_--v_--\Delta_-\bigr)\Bigr)\,.
 \label{app:eq-auxiliary-variation}
\end{align}
For $q_\lambda\neq0$, this reproduces the algebraic equations \eqref{eq:af-lambda-aux-eom}.  Together with \eqref{eq:delta-commutator}, they immediately yield \eqref{eq:af-lambda-fundamental-commutator}.

The variation with respect to $v_-$ involves $\Delta_+$, so the coefficient of $\delta v_-$ is $a_+-v_+-\Delta_+$; the $v_+$ equation follows by exchanging the $\pm$ labels. The common factor $q_\lambda$ multiplies the entire auxiliary sector.  It therefore drops out of the algebraic equations in the parameter range used here, while its sign remains essential in the action and in the sum and difference equations below.

For the group variation, write $\eta=g^{-1}\delta g$.  Useful identities include
\begin{gather}
 \delta j_\pm=\partial_\pm\eta+[j_\pm,\eta]\,,
 \qquad
\delta(g^{-1}a_-g)=[g^{-1}a_-g,\eta]\,.
\label{app:eq-group-variation-identities}
\end{gather}
The local WZW variation is $\delta S_{\mathrm{WZW}}=\kappa\int d^2x\, \tr(\eta\,\partial_+j_-)$.  In the remaining terms, it is convenient to rewrite $\tr(a_-\partial_+gg^{-1})=\tr(g^{-1}a_-g\,j_+)$. The commutator contribution from $\delta j_+$ cancels the contribution from $\delta(g^{-1}a_-g)$. Integrating by parts and using cyclicity of the trace to write the remaining terms with $\eta$ on the left, we obtain
\begin{gather}
 \delta_gS_{\mathrm{AF}\text{-}\lambda}
 =\kappa\int d^2x\,\tr(\eta\,\mathcal E_g)\,,
 \label{app:eq-group-variation}
\end{gather}
where
\begin{gather}
 \mathcal E_g=
 \partial_+\bigl(g^{-1}\partial_-g+g^{-1}a_-g\bigr)
 -\partial_-a_+
 +[a_+,g^{-1}\partial_-g+g^{-1}a_-g]\,.
 \label{app:eq-group-equation-left}
\end{gather}
Stationarity therefore sets $\mathcal E_g=0$.  The same equation in a right-invariant frame is
\begin{gather}
 \partial_-\bigl(-\partial_+gg^{-1}+ga_+g^{-1}\bigr)
 -\partial_+a_-
 +[a_-,-\partial_+gg^{-1}+ga_+g^{-1}]=0\,.
 \label{app:eq-group-equation-right}
\end{gather}
The interaction function $E$ has no explicit $g$-dependence, so it makes no direct contribution to either expression.

The equivalence of the two group equations is also a useful check on the signs.  Conjugating \eqref{app:eq-group-equation-left} by $g$, using $\partial_\pm(gXg^{-1}) =g(\partial_\pm X+[j_\pm,X])g^{-1}$, and then replacing $gj_\pm g^{-1}$ by $(\partial_\pm g)g^{-1}$ gives minus the left-hand side of \eqref{app:eq-group-equation-right}.  Thus \eqref{app:eq-group-equation-left} and \eqref{app:eq-group-equation-right} are equivalent.

Varying the gauge fields instead gives the two algebraic relations
\begin{align}
 g^{-1}\partial_-g
 -(\lambda-\Ad_g^{-1})a_-+q_\lambda v_-&=0\,,
 \notag\\
 -\partial_+gg^{-1}
 -(\lambda-\Ad_g)a_++q_\lambda v_+&=0\,.
\label{app:eq-gauge-equations}
\end{align}
For instance, the coefficient of $\delta a_+$ receives terms from the two gauged-WZW couplings, from $-\lambda\tr(a_+a_-)$, and from the linear coupling $q_\lambda\tr(a_+v_-)$.  Moving the adjoint action with invariance of the trace gives the first line of \eqref{app:eq-gauge-equations}.  The second line follows in the same way from $\delta a_-$.  This derivation is entirely algebraic and introduces no additional derivatives of the auxiliary interaction. Substituting these expressions for the Maurer-Cartan currents into the two forms of the group equation gives
\begin{align}
 \mathcal E_+={}&
 \partial_+(\lambda a_- -q_\lambda v_-)-\partial_-a_+
 +[a_+,\lambda a_- -q_\lambda v_-]=0\,,
 \notag\\
 \mathcal E_-={}&
 \partial_-(\lambda a_+ -q_\lambda v_+)-\partial_+a_-
 +[a_-,\lambda a_+ -q_\lambda v_+]=0\,.
 \label{app:eq-physical-equation-pair}
\end{align}
To form their sum, note that the commutator terms combine as
\begin{align}
 &[a_+,\lambda a_- -q_\lambda v_-]
 +[a_-,\lambda a_+ -q_\lambda v_+]
 =-q_\lambda[a_+,v_-]+q_\lambda[v_+,a_-]\doteq0\,.
 \label{app:eq-sum-commutator-cancellation}
\end{align}
Here both mixed commutators equal $[v_+,v_-]$ after the auxiliary equations are imposed.  For the difference, the same calculation instead gives
\begin{align}
 &[a_+,\lambda a_- -q_\lambda v_-]
 -[a_-,\lambda a_+ -q_\lambda v_+]
 \doteq2\lambda[a_+,a_-]-2q_\lambda[v_+,v_-]\,.
 \label{app:eq-difference-commutator}
\end{align}
The derivative terms can then be collected without further identities. After using the auxiliary commutator identity, their sum becomes
\begin{gather}
 (\lambda-1)(\partial_+a_-+\partial_-a_+)
 -q_\lambda(\partial_+v_-+\partial_-v_+)\approx0\,,
 \label{app:eq-corrected-sum-equation}
\end{gather}
whereas their difference is
\begin{align}
 &(1+\lambda)(\partial_+a_- -\partial_-a_+)
 -q_\lambda(\partial_+v_- -\partial_-v_+) + 2\lambda[a_+,a_-]-2q_\lambda[v_+,v_-]\approx0\,.
 \label{app:eq-corrected-difference-equation}
\end{align}
The flatness calculation below also checks the coefficient $2\lambda$ in \eqref{app:eq-corrected-difference-equation}.

Let us now substitute the currents defined in \eqref{eq:split-currents}. Using $q_\lambda=(\lambda-1)(\lambda+1)/\lambda$, their divergence can be written as
\begin{align}
 \partial_+\mathcal B_-+\partial_-\mathcal B_+
 ={}&-\frac{2\lambda}{(1+\lambda)(\lambda-1)}
 \Bigl[(\lambda-1)(\partial_+a_-+\partial_-a_+)
 -q_\lambda(\partial_+v_-+\partial_-v_+)\Bigr]\approx0\,.
 \label{app:eq-conserved-current-check}
\end{align}
Thus the sum equation is precisely conservation of $\mathcal B$.  For the other current, \eqref{eq:af-lambda-fundamental-commutator} gives
\begin{gather}
 [\mathcal A_+,\mathcal A_-]\doteq
 \frac{4}{(1+\lambda)^2}
 \left\{\lambda^2[a_+,a_-]+(1-\lambda^2)[v_+,v_-]\right\}\,.
 \label{app:eq-flat-current-commutator}
\end{gather}
Consequently,
\begin{align}
 \partial_+\mathcal A_- -\partial_-\mathcal A_+
 +[\mathcal A_+,\mathcal A_-] &\doteq
 \frac{2\lambda}{(1+\lambda)^2}
 \Bigl\{(1+\lambda)(\partial_+a_- -\partial_-a_+)
 -q_\lambda(\partial_+v_- -\partial_-v_+)
 \notag\\[-0.2em]
 &\hspace{7em}
 +2\lambda[a_+,a_-]-2q_\lambda[v_+,v_-]\Bigr\} \nonumber \\
 &\approx0\,.
 \label{app:eq-flat-current-check}
\end{align}
This is the flatness equation for $\mathcal A$.

The remaining identities are algebraic.  Direct expansion gives
\begin{align}
 [\mathcal B_+,\mathcal B_-]
 &\doteq\frac{4}{(1+\lambda)^2}
 \left\{\lambda^2[a_+,a_-]+(1-\lambda^2)[v_+,v_-]\right\}\,,
 \notag\\
 [\mathcal A_+,\mathcal B_-]
 &\doteq[\mathcal B_+,\mathcal A_-]
 \doteq\frac{4}{(1+\lambda)^2}
 \left\{(1+\lambda^2)[v_+,v_-]-\lambda^2[a_+,a_-]\right\}\,.
 \label{app:eq-current-commutator-comparison}
\end{align}
The first line agrees with \eqref{app:eq-flat-current-commutator}; the second line is the mixed-current identity which, in form notation, sets $[\mathcal A,\star\mathcal B]$ and $[\mathcal B,\star\mathcal A]$ to zero.  We have therefore recovered both relations in \eqref{eq:split-commutator-identities}.

To verify the form translation directly, write $\mathcal A=\mathcal A_+\dd x^++\mathcal A_-\dd x^-$ and similarly for $\mathcal B$, with $\star \left( \mathcal B \right)_\pm=\pm\mathcal B_\pm$.  Then
\begin{align}
 [\mathcal A,\mathcal A]
 &=2[\mathcal A_+,\mathcal A_-]\,\dd x^+\wedge\dd x^-\,,
 \notag\\
 [\mathcal A,\star\mathcal B]
 &=\bigl(-[\mathcal A_+,\mathcal B_-]
 +[\mathcal B_+,\mathcal A_-]\bigr)\,
 \dd x^+\wedge\dd x^-\,.
 \label{app:eq-component-form-translation}
\end{align}
The two lines of \eqref{app:eq-current-commutator-comparison} therefore give the form identities with no further equation of motion.  Similarly, $\dd(\star\mathcal B)=0$ is the component equation $\partial_+\mathcal B_-+\partial_-\mathcal B_+=0$.

Finally, it is useful to display how these four current equations enter the Lax curvature.  Expanding \eqref{eq:af-lambda-lax} before using any of the identities gives
\begin{align}
 \dd\mathfrak L+\frac12[\mathfrak L,\mathfrak L]
 = \frac{\dd\mathcal A+z\dd(\star\mathcal B)}{1-z^2} +\frac{[\mathcal A,\mathcal A]
 +2z[\mathcal A,\star\mathcal B]
 +z^2[\star\mathcal B,\star\mathcal B]}
 {2(1-z^2)^2}\,.
 \label{app:eq-expanded-lax-curvature}
\end{align}
Since $[\star\mathcal B,\star\mathcal B]=-[\mathcal B,\mathcal B]$, the algebraic relations reduce this expression to
\begin{gather}
 \dd\mathfrak L+\frac12[\mathfrak L,\mathfrak L]
 \doteq\frac{1}{1-z^2}
 \left\{\dd\mathcal A+\frac12[\mathcal A,\mathcal A]
 +z\dd(\star\mathcal B)\right\}\,.
 \label{app:eq-lax-curvature-reduction}
\end{gather}
The sum and difference equations now set the two terms in braces to zero for every value of the spectral parameter.

In light-cone components, the same statement takes the particularly simple form
\begin{align}
 \mathcal F_{+-}(z)
 \doteq\frac{1}{1-z^2}\Bigl(&
 \partial_+\mathcal A_- -\partial_-\mathcal A_+
 +[\mathcal A_+,\mathcal A_-] - z(\partial_+\mathcal B_-+\partial_-\mathcal B_+)\Bigr)\approx0\,.
 \label{app:eq-lax-curvature-components}
\end{align}
After using the algebraic compatibility relations and canceling a common factor of $1-z^2$, the numerator is linear in $z$. Its constant and linear coefficients give flatness of $\mathcal A$ and conservation of $\mathcal B$, respectively.

\section{Hamiltonian constraints and WZW currents}
\label{app:hamiltonian-constraints}

We now justify the Hamiltonian assumption used in Subsection~\ref{subsec:current-algebra}.  The relevant observation is a simple block-matrix property of auxiliary constraints; this feature has been discussed in other examples of auxiliary field sigma models in several previous works \cite{Bielli:2024ach,Cesaro:2024ipq,Bielli:2026ggs}, but we review it here to make the present article self-contained. We then apply this result to the fields $a_\pm,v_\pm$ and explain why the canonical WZW currents may be represented by expressions containing those fields after the brackets have been computed.

Consider a phase space with canonical variables collectively denoted by $(Z,\Pi)$, enlarged by nondynamical coordinates $y^I$ and their momenta $p_I$.  Their primary constraints are $\phi_I=p_I\simeq0$.  Preserving these in time produces secondary constraints $\psi_I(Z,\Pi;y)\simeq0$.  We suppose that the Dirac algorithm closes at this stage and define
\begin{gather}
 M_{IJ}=\{\psi_I,\phi_J\}
       =\frac{\delta\psi_I}{\delta y^J}\,,
 \qquad
 N_{IJ}=\{\psi_I,\psi_J\}\,.
 \label{app:eq-regularity-block}
\end{gather}
If $M$ is invertible, the constraint matrix for $\Omega=(\phi,\psi)$ and its inverse are
\begin{gather}
 D=
 \begin{pmatrix}
 0&-M^T\\ M&N
 \end{pmatrix}\,,
 \qquad
 D^{-1}=
 \begin{pmatrix}
 M^{-1}NM^{-T}&M^{-1}\\ -M^{-T}&0
 \end{pmatrix}\,.
 \label{app:eq-constraint-block-inverse}
\end{gather}
Here the collective indices include Lie-algebra and spatial labels, so matrix multiplication includes the corresponding integrations.  The vanishing lower-right block is the important point.  It follows directly from block multiplication and does not require the secondary constraints to commute among themselves.

Invertibility of $M$ also ensures that the constraint algorithm closes at this stage.  If the total Hamiltonian is $H_T=H+u^I\phi_I$, preservation of the secondary constraints gives
\begin{gather}
 \dot\psi_I\simeq\{\psi_I,H\}+M_{IJ}u^J=0\,.
 \label{app:eq-secondary-preservation}
\end{gather}
Since $M$ is invertible, this equation fixes the multipliers $u^I$, so no further constraint arises in this sector. The same invertibility assumption therefore gives the block inverse and ensures that the auxiliary constraint algorithm closes.

To see the consequence for observables, recall that the Dirac bracket is
\begin{gather}
 \{F,G\}_D=\{F,G\}
 -\{F,\Omega_{\mathcal I}\}
 (D^{-1})^{\mathcal I\mathcal J}
 \{\Omega_{\mathcal J},G\}\,.
 \label{app:eq-dirac-bracket-definition}
\end{gather}
If $F(Z,\Pi)$ and $G(Z,\Pi)$ are independent of $y,p$, then the primary-constraint terms in their Dirac bracket vanish, while the secondary-secondary term is multiplied by that zero block.  Therefore
\begin{gather}
 {\ \{F(Z,\Pi),G(Z,\Pi)\}_D
       =\{F(Z,\Pi),G(Z,\Pi)\}\,.\ }
 \label{app:eq-wzw-observable-bracket}
\end{gather}
Notice that $F$ and $G$ need not commute with the secondary constraints. Their possible brackets with $\psi_I$ occur only in the secondary-secondary correction, which vanishes because of the lower-right block of $D^{-1}$.  Thus eliminating the auxiliary variables leaves the Poisson brackets of observables $F(Z,\Pi)$ and $G(Z,\Pi)$ unchanged. This familiar auxiliary field argument will be applied in the gauge-fixed formulation used here; see, for instance, \cite{Dirac:1964:LQM,henneaux1992quantization} for the general constrained Hamiltonian framework.

The order of these operations matters.  Before gauge fixing, the parent construction of the lambda model \cite{Sfetsos:2013wia} contains a first-class gauge symmetry, and the matrix of all constraints cannot be inverted without accompanying gauge conditions.  The action used here has already fixed this gauge symmetry by setting the PCM field to the identity. We apply the block argument to the remaining nondynamical fields.  No velocities of $a_\pm$ or $v_\pm$ occur in \eqref{eq:af-lambda-action}, so their vanishing momenta are manifest in this formulation.

For the present theory we take
\begin{gather}
 y^I=(a_+^A,a_-^A,v_+^A,v_-^A)\,.
 \label{app:eq-algebraic-variables}
\end{gather}
Their conjugate momenta vanish, and a convenient basis for the secondary constraints is
\begin{align}
 \Psi^a_+&=\mathscr J_+
 -\kappa(\lambda a_+-q_\lambda v_+-a_-)\simeq0\,,
 \notag\\
 \Psi^a_-&=\mathscr J_-
 -\kappa(\lambda a_--q_\lambda v_--a_+)\simeq0\,,
 \notag\\
 \Psi^v_+&=a_+-v_+-\Delta_+\simeq0\,,
 \notag\\
 \Psi^v_-&=a_--v_--\Delta_-\simeq0\,.
 \label{app:eq-model-secondary-constraints}
\end{align}
The currents $\mathscr J_\pm$ in this display are canonical functions of the WZW coordinates and momenta.  The first pair of equations comes from the gauge fields, and the second pair is the auxiliary equation.

The canonical WZW currents generate left and right transformations of $g$ through Poisson brackets. When the gauge field equations are written in Hamiltonian variables, the WZW momentum appears only through these currents; the remaining terms are algebraic in $a_\pm,v_\pm$. Changing the basis of secondary constraints leaves the Dirac bracket unchanged. We use \eqref{app:eq-model-secondary-constraints} to apply the block-matrix argument above and relate the WZW currents to $a_\pm,v_\pm$.

To make the required invertibility condition explicit, define the local auxiliary Hessian by
\begin{gather}
 \mathcal H_{st}=\frac{\delta\Delta_s}{\delta v_t}\,,
 \qquad s,t\in\{+,-\}\,.
 \label{app:eq-auxiliary-hessian}
\end{gather}
Suppressing its adjoint indices and spatial delta function, the Jacobian of \eqref{app:eq-model-secondary-constraints} with respect to the four fields in \eqref{app:eq-algebraic-variables} is
\begin{gather}
 M=
 \begin{pmatrix}
 -\kappa\lambda&\kappa&\kappa q_\lambda&0\\
 \kappa&-\kappa\lambda&0&\kappa q_\lambda\\
 1&0&-1-\mathcal H_{++}&-\mathcal H_{+-}\\
 0&1&-\mathcal H_{-+}&-1-\mathcal H_{--}
 \end{pmatrix}\,.
 \label{app:eq-model-regularity-matrix}
\end{gather}
Every scalar entry in the upper two rows multiplies the identity in adjoint space.  The Hessian entries in the lower rows may mix the two light-cone labels and the adjoint components, but for the interactions considered here they are local in the spatial coordinate.  Thus $M$ is a pointwise finite-dimensional matrix tensored with the appropriate delta kernel.  The form \eqref{app:eq-model-regularity-matrix} is also useful beyond perturbation theory: for any chosen interaction and configuration it gives a concrete test for whether the algebraic constraints can be eliminated. At $E=0$, the Hessian vanishes.  The lower-right block is then $-\mathbf 1$, and its Schur complement is
\begin{gather}
 \kappa
 \begin{pmatrix}
 -\lambda&1\\[0.1em]1&-\lambda
 \end{pmatrix}
 +\kappa q_\lambda\mathbf 1
 =\kappa
 \begin{pmatrix}
 -\lambda^{-1}&1\\[0.1em]1&-\lambda^{-1}
 \end{pmatrix}\,.
 \label{app:eq-seed-schur-complement}
\end{gather}
Thus, for each adjoint component,
\begin{gather}
 \det M\big|_{E=0}
 =\kappa^2\left(\lambda^{-2}-1\right)\,.
 \label{app:eq-seed-regularity-check}
\end{gather}
This is nonzero for $0<\lambda<1$ and $\kappa\neq0$, and remains so for sufficiently small continuous deformations. For a general interaction function, one must check whether the matrix in \eqref{app:eq-model-regularity-matrix} is invertible.  If it loses rank, the Dirac algorithm must be reconsidered.

We finish by clarifying the relation between canonical and Lagrangian currents.  The canonical WZW currents $\mathscr J_\pm(g,\Pi_g)$ are independent of $a_\pm,v_\pm$, so the block argument gives them the standard Kac-Moody brackets quoted in \eqref{eq:wzw-km-algebra}. Their Lagrangian representatives were given in (\ref{canonical_wzw_currents}), which we repeat for convenience:
\begin{align}
 \mathscr J_+
 &=\kappa\bigl(-\partial_+gg^{-1}+ga_+g^{-1}-a_-\bigr)\,,
 \notag\\
 \mathscr J_-
 &=\kappa\bigl(g^{-1}\partial_-g+g^{-1}a_-g-a_+\bigr)\,.
 \label{app:eq-lagrangian-wzw-representatives}
\end{align}
These expressions follow by substituting the formula for $\Pi_g$ in terms of the fields and velocities.  One must therefore compute the brackets of the canonical currents first and impose \eqref{app:eq-model-secondary-constraints} afterward.  The latter step gives the relations in \eqref{eq:wzw-reduced-constraints} and hence the linear transformation \eqref{eq:wzw-split-current-relation}.  Since this transformation is invertible for the parameter range used in the Hamiltonian analysis, the current algebra in \eqref{eq:split-current-algebra} inherits the Jacobi identities from the two WZW algebras.

One can verify the last statement without introducing any new Poisson brackets.  Adding and subtracting the two relations for $\mathscr J_\pm$ solves linearly for $\mathcal B_\tau$ and $\mathcal A_\sigma$, as shown in \eqref{eq:wzw-split-current-inverse}.  Substitution into the two Kac-Moody algebras then gives all three lines of \eqref{eq:split-current-algebra}.  The auxiliary interaction affects the Hamiltonian and the solution of the algebraic equations, but it does not appear in this linear transformation.  This is the reason that the coefficients $m_i$ used in the Yangian and Maillet calculations contain $\lambda$ and $\kappa$, but no derivatives of $E$.

\section{Yangian regulator and Maillet matching}
\label{app:yangian-maillet}

This appendix gives details of two calculations used in Subsections~\ref{subsec:yangian} and~\ref{subsec:maillet}. First, we explain the ordered endpoint prescription for the nonlocal Yangian charges. Second, we compare the scalar coefficients directly to put the spatial Lax bracket into Maillet form and fix the normalization of the twist function.

Let $\sigma_i>0$, for $i=1,\ldots,6$, be six independent cutoff parameters. At finite values of these cutoffs, define the potential
\begin{gather}
 \chi_{\boldsymbol\sigma}^{(0)}(x)
 =\beta_{\mathrm Y}\int_{-\sigma_1}^{+\sigma_2}d y\,
 \theta(x-y)\mathcal B_\tau(y)\,.
 \label{app:eq-yangian-regulated-potential}
\end{gather}
The level-zero charge and the local and bilocal parts of the level-one charge are then regulated as
\begin{align}
 Q_{\boldsymbol\sigma}^{(0)A}
 &=\beta_{\mathrm Y}\int_{-\sigma_1}^{+\sigma_2}d x\,
 \mathcal B_\tau^A(x)\,,
 \notag\\
 Q_{\boldsymbol\sigma}^{(1)A}
 &=\int_{-\sigma_5}^{+\sigma_6}d x\,
 \mathcal A_\sigma^A(x)
 +\frac{1}{2\beta_{\mathrm Y}}
 \int_{-\sigma_3}^{+\sigma_4}d x\,
 [\mathcal B_\tau(x),\chi_{\boldsymbol\sigma}^{(0)}(x)]^A\,.
 \label{app:eq-yangian-regulated-charges}
\end{align}
All six cutoffs are sent to infinity while preserving the strict order
\begin{gather}
 \sigma_6<\sigma_4<\sigma_2\,,
 \qquad
 \sigma_5<\sigma_3<\sigma_1\,.
 \label{eq:yangian-regulator-order}
\end{gather}
This is the endpoint prescription of the Yangian theorem in \cite{Bielli:2026ggs}, following the analyses of \cite{MacKay:1992he,Klose:2016qfv}.  The two chains order the positive and negative endpoints independently. Integrating the differentiated delta function over the larger interval first produces endpoint distributions outside the remaining integration domain. Bulk terms obtained by differentiating Heaviside factors are retained and treated in the Serre calculation.  With sufficient falloff, $\chi_{\boldsymbol\sigma}^{(0)}$ approaches the base-point potential in \eqref{eq:yangian-current-densities}, and the regulated charges approach \eqref{eq:yangian-charges}.

Let us make the role of the strict inequalities more explicit.  On a finite interval, integration against a differentiated delta function gives
\begin{align}
 \int_a^b d x\,F(x)\partial_x\delta(x-y)
 ={}&F(b)\delta(b-y)-F(a)\delta(a-y) -\int_a^b d x\,\partial_xF(x)\delta(x-y)\,.
 \label{app:eq-finite-interval-delta-prime}
\end{align}
The two boundary terms vanish only after checking that their support at $y=a$ and $y=b$ lies outside the domain of the remaining $y$-integration. At the positive end, the local part of $Q^{(1)}$ stops at $\sigma_6$, the bilocal part at $\sigma_4$, and the potential at $\sigma_2$.  With the order $\sigma_6<\sigma_4<\sigma_2$, integrating over the larger interval first removes successive endpoint distributions before taking any limit.  The negative endpoints work in the same way with $\sigma_5<\sigma_3<\sigma_1$.  Once the Poisson brackets have been evaluated in this order, all six parameters may be taken to infinity. Inside the integration region,
\begin{gather}
 \partial_x\chi_{\boldsymbol\sigma}^{(0)}(x)
 =\beta_{\mathrm Y}\mathcal B_\tau(x)\,,
 \label{app:eq-regulated-potential-derivative}
\end{gather}
so $\chi_{\boldsymbol\sigma}^{(0)}$ approaches the potential entering the generalized BIZZ current. A common cutoff would instead require an additional prescription for the coincident endpoints.

The local term in $Q^{(1)}$ and the bilocal term play complementary roles in the charge algebra.  Bracketing the level-zero density with either term produces the adjoint action, while the derivative of the delta function in the mixed current bracket generates precisely the endpoint contributions in \eqref{app:eq-finite-interval-delta-prime}.  After the ordered limit, these terms vanish and the normalization $\beta_{\mathrm Y}=m_1^{-1}$ converts the remaining bulk expression into the standard Yangian bracket. The same prescription is used to compute the bracket of two level-one charges and obtain the Serre relation in the main text.

We now turn to the Maillet calculation.  Set
\begin{gather}
 D(z,z')=(1-z^2)(1-z'^2)\,.
 \label{app:eq-maillet-denominator}
\end{gather}
Using the current algebra \eqref{eq:split-current-algebra} to evaluate the bracket of the spatial Lax matrix \eqref{eq:af-lambda-spatial-lax} gives
\begin{align}
 \{\mathfrak L_\sigma^A(x,z),
    \mathfrak L_\sigma^B(y,z')\}_D
 ={}&\frac{f^{AB}{}_{C}}{D(z,z')}
 \left\{m_1(z+z')\mathcal A_\sigma^C
 +(m_6+m_1zz')\mathcal B_\tau^C\right\}(x)\delta(x-y)
 \notag\\
 &+\frac{m_4(z+z')}{D(z,z')}
 \gamma^{AB}\partial_x\delta(x-y)\,.
 \label{app:eq-component-lax-bracket}
\end{align}
The vanishing coefficients in \eqref{eq:af-lambda-mi} remove the other possible tensor structures.

To obtain \eqref{app:eq-component-lax-bracket}, one expands $\mathfrak L_\sigma=(\mathcal A_\sigma+z\mathcal B_\tau)/(1-z^2)$ in both slots of the bracket.  The two mixed brackets supply the terms linear in $z$ and $z'$, while the $\{\mathcal B_\tau,\mathcal B_\tau\}_D$ bracket supplies the coefficient $m_1zz'$.  Because $m_2=m_5=m_7=0$ and $m_3=m_1$, the result can involve only the two fields displayed in the first line.  The central term comes from the mixed bracket and is proportional to $m_4(z+z')$.  This direct component calculation fixes the relative sign between the ultralocal and non-ultralocal terms before any Maillet ansatz is made.

In the convention of \eqref{eq:af-lambda-maillet-bracket}, take the ansatz $r_{12}(z,z')=f(z,z')C_{12}$, with $f$ independent of the fields.  Comparing the coefficients of $[\mathcal A_{\sigma,2},C_{12}]$ and $[\mathcal B_{\tau,2},C_{12}]$ gives the two scalar equations
\begin{align}
 \frac{f(z,z')}{1-z^2}
 +\frac{f(z',z)}{1-z'^2}
 &=\frac{m_1(z+z')}{D(z,z')}\,,
 \notag\\
 \frac{zf(z,z')}{1-z^2}
 +\frac{z'f(z',z)}{1-z'^2}
 &=\frac{m_6+m_1zz'}{D(z,z')}\,.
 \label{app:eq-maillet-scalar-equations}
\end{align}
They are solved simultaneously by
\begin{gather}
 f(z,z')=\frac{m_6-m_1z'^2}{(1-z'^2)(z-z')}\,.
\label{app:eq-maillet-f-solution}
\end{gather}
Indeed, for the first equation the common numerator is
\begin{align}
 (m_6-m_1z'^2)-(m_6-m_1z^2) =m_1(z-z')(z+z')\,,
 \label{app:eq-maillet-first-numerator}
\end{align}
whereas multiplication by $z$ and $z'$ gives
\begin{align}
 z(m_6-m_1z'^2)-z'(m_6-m_1z^2) =(z-z')(m_6+m_1zz')\,.
 \label{app:eq-maillet-second-numerator}
\end{align}
The factor $z-z'$ cancels the pole in \eqref{app:eq-maillet-f-solution}, reproducing both ultralocal coefficients in \eqref{app:eq-component-lax-bracket}.  The derivative term provides the remaining sign check.  The symmetric part of the kernel is
\begin{gather}
 f(z,z')+f(z',z)
 =\frac{(m_1-m_6)(z+z')}{D(z,z')}
 =-\frac{m_4(z+z')}{D(z,z')}\,,
 \label{app:eq-maillet-symmetric-part}
\end{gather}
where the last equality uses $m_4=m_6-m_1$.  Since the convention in \eqref{eq:af-lambda-maillet-bracket} contains $-s_{12}\partial_x\delta(x-y)$, this reproduces the coefficient $+m_4(z+z')/D(z,z')$ in \eqref{app:eq-component-lax-bracket}.  The scalar comparison therefore yields the kernels displayed in \eqref{eq:af-lambda-rs}.

The conversion between the component and tensor expressions uses only invariance of the quadratic Casimir:
\begin{gather}
 f^{AB}{}_{C}X^CT_A\otimes T_B=[X_2,C_{12}]\,,
 \qquad
 [C_{12},X_1]=-[C_{12},X_2]\,.
 \label{app:eq-casimir-tensor-identities}
\end{gather}
Thus no field-dependent tensor is needed.  In particular, the resulting Maillet bracket has kernels proportional to $C_{12}$ and inherits the Jacobi identity from the Dirac bracket calculated in Appendix~\ref{app:hamiltonian-constraints}.

The same solution immediately identifies the twist function.  Writing $f(z,z')=(z-z')^{-1}\varphi_\lambda^{-1}(z')$ gives
\begin{gather}
 \varphi_\lambda^{-1}(z)
 =\frac{m_6-m_1z^2}{1-z^2}
 =m_1\,\frac{z_\lambda^2-z^2}{1-z^2}\,.
 \label{app:eq-inverse-twist-check}
\end{gather}
Since $m_1^{-1}=\kappa(1-\lambda^2)/\lambda$, inversion reproduces \eqref{eq:af-lambda-twist}, including its overall normalization.  In this form it is also clear that the zeros of the inverse twist function occur at $z=\pm z_\lambda$.

We close with two short checks of the twist normalization.  The constants in \eqref{eq:wzw-split-current-relation} obey
\begin{gather}
 \frac{c_\lambda}{d_\lambda}
 =z_\lambda=\frac{1-\lambda}{1+\lambda}\,,
 \qquad
 d_\lambda(1-z_\lambda^2)=2\kappa\,.
\label{app:eq-twist-site-normalization}
\end{gather}
The poles of the twist function also follow from the current-algebra coefficients:
\begin{gather}
 \frac{m_6}{m_1}
 =\frac{(1-\lambda)^2}{(1+\lambda)^2}
 =z_\lambda^2\,.
 \label{app:eq-twist-ratio-check}
\end{gather}
This relation gives the factor $z_\lambda^2-z^2$ in the denominator of $\varphi_\lambda(z)$ in \eqref{eq:af-lambda-twist}. Substituting $z=\pm z_\lambda$ into the spatial Lax matrix gives
\begin{gather}
 \mathfrak L_\sigma(z_\lambda)
 =\frac{\mathcal A_\sigma+z_\lambda\mathcal B_\tau}
        {1-z_\lambda^2}
 =\frac{\mathscr J_+}{2\kappa}\,,
 \qquad
 \mathfrak L_\sigma(-z_\lambda)
 =-\frac{\mathscr J_-}{2\kappa}\,.
 \label{app:eq-lax-twist-site-check}
\end{gather}
For the twist function \eqref{eq:af-lambda-twist}, the identity $(1-z_\lambda^2)/z_\lambda=4\lambda/(1-\lambda^2)$ similarly gives
\begin{gather}
 \operatorname*{res}_{z=z_\lambda}\varphi_\lambda(z)\,d z
 =-2\kappa\,,
 \qquad
 \operatorname*{res}_{z=-z_\lambda}\varphi_\lambda(z)\,d z
 =+2\kappa\,.
\label{app:eq-twist-residue-check}
\end{gather}
For instance, the residue at $z=z_\lambda$ is
\begin{gather}
 \frac{\kappa(1-\lambda^2)}{\lambda}
 \frac{1-z_\lambda^2}{-2z_\lambda}
 =-2\kappa\,,
 \label{app:eq-positive-twist-residue}
\end{gather}
and the sign reverses at $z=-z_\lambda$.  This arithmetic checks the overall prefactor in the twist function, which is fixed by the affine levels and cannot be discarded independently of the normalization of $C_{12}$. The two evaluations and the two residues reproduce the normalizations and opposite affine levels of the WZW currents.

\bibliographystyle{utphys}
\bibliography{master}

\end{document}